\documentclass[prl,twocolumn,superscriptaddress]{revtex4-2}
\usepackage{}
\usepackage{amssymb}
\usepackage{amsfonts}
\usepackage{bbm}
\usepackage{mathrsfs}
\usepackage{graphics,graphicx,epsfig,bm,amsmath,amsthm,amssymb}
\usepackage{bm}
\usepackage{bbm}
\usepackage{longtable}
\usepackage{multirow}
\usepackage{array}
\usepackage{color}
\usepackage{diagbox}
\usepackage{amsmath}
\usepackage{upgreek}
\usepackage{multirow}
\usepackage{threeparttable}
\usepackage{braket}
\usepackage[usenames,dvipsnames]{xcolor}

\usepackage{float}
\usepackage[a4paper,colorlinks=true,
linkcolor=blue,citecolor=blue,
pdfauthor={ },
pdftitle={ },
pdfsubject={ },
pdfkeywords={ }]{hyperref}

\newcommand{\tabincell}[2]{\begin{tabular}{@{}#1@{}}#2\end{tabular}}
\begin{document}

	\title{Third-order exceptional line in a nitrogen-vacancy spin system}
	\affiliation{CAS Key Laboratory of Microscale Magnetic Resonance and School of Physical Sciences, University of Science and Technology of China, Hefei 230026, China}
	\affiliation{CAS Center for Excellence in Quantum Information and Quantum Physics, University of Science and Technology of China, Hefei 230026, China}
	\affiliation{School of Science, Beijing University of Posts and Telecommunications, Beijing 100876, China}
	\affiliation{Hefei National Laboratory, University of Science and Technology of China, Hefei 230088, China}
	\affiliation{Institute of Quantum Sensing and School of Physics, Zhejiang University, Hangzhou 310027, China}

	\author{Yang Wu}
	\thanks{These authors contributed equally to this work.}
	\affiliation{CAS Key Laboratory of Microscale Magnetic Resonance and School of Physical Sciences, University of Science and Technology of China, Hefei 230026, China}
	\affiliation{CAS Center for Excellence in Quantum Information and Quantum Physics, University of Science and Technology of China, Hefei 230026, China}

	\author{Yunhan Wang}
	\thanks{These authors contributed equally to this work.}
	\affiliation{CAS Key Laboratory of Microscale Magnetic Resonance and School of Physical Sciences, University of Science and Technology of China, Hefei 230026, China}
	\affiliation{CAS Center for Excellence in Quantum Information and Quantum Physics, University of Science and Technology of China, Hefei 230026, China}

	\author{Xiangyu Ye}
	\thanks{These authors contributed equally to this work.}
	\affiliation{CAS Key Laboratory of Microscale Magnetic Resonance and School of Physical Sciences, University of Science and Technology of China, Hefei 230026, China}
	\affiliation{CAS Center for Excellence in Quantum Information and Quantum Physics, University of Science and Technology of China, Hefei 230026, China}
	
	\author{Wenquan Liu}
	\affiliation{CAS Key Laboratory of Microscale Magnetic Resonance and School of Physical Sciences, University of Science and Technology of China, Hefei 230026, China}
	\affiliation{CAS Center for Excellence in Quantum Information and Quantum Physics, University of Science and Technology of China, Hefei 230026, China}
	\affiliation{School of Science, Beijing University of Posts and Telecommunications, Beijing 100876, China}
	
	\author{Zhibo Niu}
	\affiliation{CAS Key Laboratory of Microscale Magnetic Resonance and School of Physical Sciences, University of Science and Technology of China, Hefei 230026, China}
	\affiliation{CAS Center for Excellence in Quantum Information and Quantum Physics, University of Science and Technology of China, Hefei 230026, China}
	
	\author{Chang-Kui Duan}
	\affiliation{CAS Key Laboratory of Microscale Magnetic Resonance and School of Physical Sciences, University of Science and Technology of China, Hefei 230026, China}
	\affiliation{CAS Center for Excellence in Quantum Information and Quantum Physics, University of Science and Technology of China, Hefei 230026, China}
	\affiliation{Hefei National Laboratory, University of Science and Technology of China, Hefei 230088, China}
	
	\author{Ya Wang}
	\affiliation{CAS Key Laboratory of Microscale Magnetic Resonance and School of Physical Sciences, University of Science and Technology of China, Hefei 230026, China}
	\affiliation{CAS Center for Excellence in Quantum Information and Quantum Physics, University of Science and Technology of China, Hefei 230026, China}
	\affiliation{Hefei National Laboratory, University of Science and Technology of China, Hefei 230088, China}

	\author{Xing Rong}
	\email{xrong@ustc.edu.cn}
	\affiliation{CAS Key Laboratory of Microscale Magnetic Resonance and School of Physical Sciences, University of Science and Technology of China, Hefei 230026, China}
	\affiliation{CAS Center for Excellence in Quantum Information and Quantum Physics, University of Science and Technology of China, Hefei 230026, China}
	\affiliation{Hefei National Laboratory, University of Science and Technology of China, Hefei 230088, China}

	\author{Jiangfeng Du}
	\email{djf@ustc.edu.cn}
	\affiliation{CAS Key Laboratory of Microscale Magnetic Resonance and School of Physical Sciences, University of Science and Technology of China, Hefei 230026, China}
	\affiliation{CAS Center for Excellence in Quantum Information and Quantum Physics, University of Science and Technology of China, Hefei 230026, China}
	\affiliation{Hefei National Laboratory, University of Science and Technology of China, Hefei 230088, China}
	\affiliation{Institute of Quantum Sensing and School of Physics, Zhejiang University, Hangzhou 310027, China}

	\begin{abstract}
		
		The exceptional points (EPs) aroused from the non-Hermiticity bring rich phenomena, such as exceptional nodal topologies\cite{PRB_Stalhammar}, unidirectional invisibility\cite{NPho_Chang,NP_Peng,PRL_Lin_2011}, single-mode lasing\cite{Sci_Feng,Sci_Hodaei}, sensitivity enhancement\cite{Nat_Chen,Nat_Hokmabadi} and energy harvesting\cite{ComPhy_FA}.
		Isolated high-order EPs have been observed \cite{Nat_Hodaei,Nat_Patil,Sci_Tang} to exhibit richer topological characteristics \cite{PRX_Ding,PRL_Deplace,PRL_Mandal} and better performance in sensing \cite{NJP_Zeng,PRApp_Wang,OE_Zeng,Nat_Hodaei} over 2nd-order EPs.
		Recently, high-order EP geometries, such as lines or rings formed entirely by high-order EPs, are predicted to provide richer phenomena and advantages over stand-alone high-order EPs\cite{Nat_Rev_Phys}.
		However, experimental exploration of high-order EP geometries is hitherto beyond reach due to the demand of more degrees of freedom in the Hamiltonian's parameter space or a higher level of symmetries.
		Here we report the observation of the third-order exceptional line (EL) at the atomic scale.
		By introducing multiple symmetries, the emergence of the third-order EL has been successfully realized with a single electron spin of nitrogen-vacancy center in diamond.
		Furthermore, the behaviors of the EP structure under different symmetries are systematically investigated.
		The symmetries are shown to play essential roles in the occurrence of high-order EPs and the related EP geometries.
		Our work opens a new avenue to explore high-order EP-related topological physics at the atomic scale\cite{PRL_Deplace,PRL_Mandal} and to the potential applications of high-order EPs in quantum technologies\cite{NC_Lau,PR_Wiersig,PRL_Yu}.
		
	\end{abstract}

	\maketitle

	Hermiticity lies at the heart of quantum mechanics, ensuring the probability conservation and the real-valuedness of the energy spectrum. However, the presence of flows of energy, particles and information outside the Hilbert space we focus on is ubiquitous in nature, where non-Hermitian physics debuts. Non-Hermitian physics has recently been extensively studied in quantum systems, revealing exotic topological phenomena as well as novel dynamical features\cite{RMP_Bergholtz,PRL_NHtopo1,PRX_NHtopo1,PRX_NHtopo2,Science_NHtopo,PRL_NHtopo2,PRA_NHtopo,NHedge,NHBB,encircle,statecontrol1,statecontrol2,statecontrol3}. The non-Hermitian singularities, where $k~( k\ge2)$ eigenvalues and eigenstates coalesce, are called $k$th-order exceptional points (EP$k$s) and they play an essential role in understanding the topological phase transitions\cite{phasetransition} and topological state control\cite{encircle}. Great efforts have been made in exploring experimentally the second-order EPs in quantum systems, including the demonstration of the topological phase with half-integer winding number\cite{NHtopo} due to the encircling of an EP, new routes of quantum state control using the dynamical properties in the vicinity of an EP\cite{PRL_Liu,PRL_Abbasi}, and observation of the spontaneous $\mathcal{PT}$ symmetry broken through an EP\cite{Sci_Wu,epion}.

	Recently, high-order exceptional points have been proposed to host features beyond EP2s with respect to topological properties and enhanced sensing performance. The emergence of high-order EPs usually accompanied by very complicated energy surface structures which have abundant topological features\cite{PRX_Ding,PRL_Deplace}, and the dispersion relation near the high-order EPs also has advantage EP2s in the enhancement of sensitivity\cite{Nat_Hodaei,NJP_Zeng,PRApp_Wang,OE_Zeng}. There are experiments for investigating the high-order EPs in the optics\cite{Nat_Hodaei}, cavity optomechanics\cite{Nat_Patil} and acoustics\cite{Sci_Tang} systems. These works focused on physics related to isolated high-order EPs.  High-order EP geometries, such as lines or rings formed entirely by high-order EPs, are expected to provide advantages over stand-alone high-order EPs\cite{Nat_Rev_Phys}.
	Since more degrees of freedom in the Hamiltonian's parameter space or additional more symmetries are required \cite{PRL_Deplace}, experimental investigation on the high-order EP geometries is challenging and remains elusive.
	In this work, we report the experimental observation of third-order exceptional line (EL) at the atomic scale utilizing a single nitrogen-vacancy (NV) center in diamond.
	By introducing symmetries, the third-order EL was successfully observed experimentally to emerge in two-dimensional parameter space.
	Furthermore, we also systemically investigated the role of symmetry in the high-order EP geometries.
	
\begin{figure}[http]
		\centering
		\includegraphics[width=1\columnwidth]{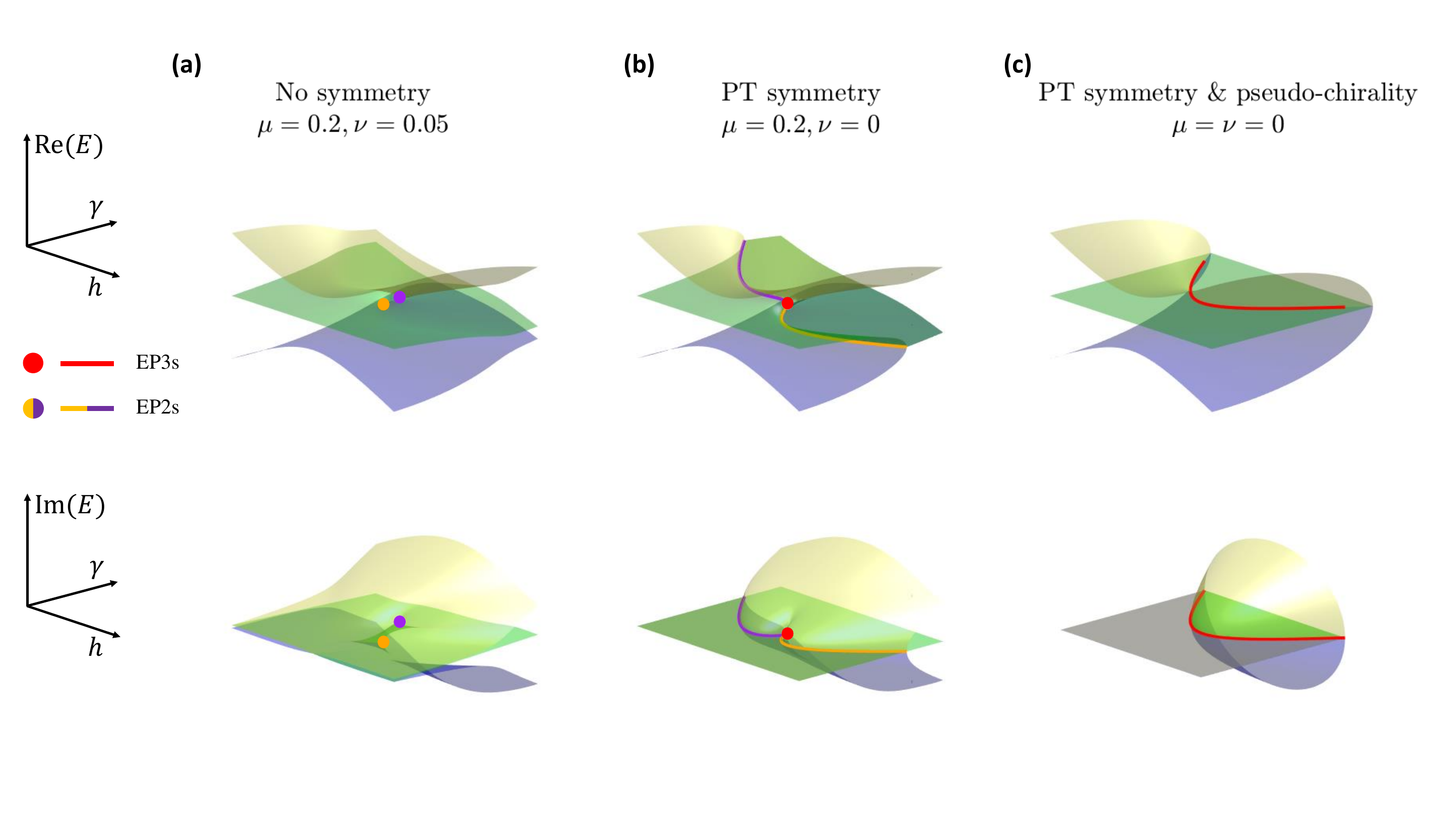}
		\caption{ The third-order exceptional point structures of $H^{(\mu,\nu)}(\gamma,h)$ with different symmetries.
			(a-c) The colored surfaces are the eigenvalue Riemann sheets of $H^{(\mu,\nu)}(\gamma,h)$ for the real (upper panel) and imaginary (lower panel) parts with different symmetries. Yellow, green and purple sheets correspond to eigenvalues $E_1$, $E_2$ and $E_3$, respectively. (a) Without any symmetry, only isolated EP2s (purple and orange points) exit without any EP3. (b) Introducing $\mathcal{PT}$ symmetry, isolated EP3 (red point) appears at the cusp of two second-order lines (purple and orange lines). (c) The third-order EL (red line) emerges when both $\mathcal{PT}$ symmetry and pseudo-chirality are introduced.
		}
		\label{Fig1}
	\end{figure}

	We consider the non-Hermitian Hamiltonian of the following form
	\begin{equation}
		H^{(\mu,\nu)}(\gamma,h)=S_x+(i\gamma+\nu) S_z+hS_y+\mu\left( \begin{array}{ccc}
			0  & 1 & 0\\
			-1 & 0 & -1\\
			0  & 1 & 0
		\end{array}
		\right ),
		\label{Hamiltonian}
	\end{equation}
	where $S_x$, $S_y$ and $S_z$ are spin-1 operators, $\gamma$ and $h$ are two real-valued variables of the non-Hermitian Hamiltonian, while real-valued $\nu$ and $\mu$ characterize the perturbations which control the symmetry of the Hamiltonian.
	It is worth noting that the EP2s are stable in the systems with two free parameters\cite{PRL_YangZs}.
	This is, however, not the case for high-order EPs.
	The existence of the EP3s requires that the characteristic polynomial $P(E)\equiv \mathrm{det}[H^{(\mu,\nu)}(\gamma,h)-E]=f_3E^3+f_2E^2+f_1E+f_0$ has threefold multiple roots, where $f_3=1$, $f_2=0$, $f_1 = \gamma^2-1-h^2-2i\gamma \nu-\nu^2+2\mu^2$, $f_0 = -2\sqrt{2}\mu h(i\nu-\gamma)$.
	This constraint is demonstrated to be equivalent to the resultants $R_{P,P^{'}}$ and $R_{P^{'},P{''}}$ to vanish\cite{PRL_Deplace}, where $P^{'}$ and $P{''}$ are the first and second derivatives of the polynomial $P(E)$, respectively (See Supplementary Materials, S1 for the definition of the resultants).
	The explicit expressions of the resultants for the Hamiltonian $H^{(\mu,\nu)}(\gamma,h)$ are $R_{P,P'}=4f_1^3+27f_0^2$ and $R_{P',P''}=36f_0$.
	The condition $\text{Re}(R_{P,P'})=\text{Im}(R_{P,P'})=\text{Re}(R_{P',P''})=\text{Im}(R_{P',P''})=0$ requires that a non-Hermitian system without symmetry protection needs four free real parameters for the possible existence of EP3 (See Supplementary Materials, S1 for the derivation).
	When $\mu$ and $\nu$ in Eq.~\ref{Hamiltonian} are both nonzero constants, there is no symmetry protection in the non-Hermitian Hamiltonian.
	The $H^{(\mu,\nu)}(\gamma,h)$ contains only two free parameters $\gamma$ and $h$ and so cannot satisfy the four constraints at the same time.
	Thus, as shown in Fig.~\ref{Fig1}a, no EP3 can be found in the $\gamma-h$ plane, while EP2s (purple and orange points) exist.
	
	When symmetries are introduced to reduce the number of constraints, EP3s and the more complicated geometries of EP3 can appear in two-dimensional parameter space.
	For example, when we introduce the symmetry constraints such as parity-time symmetry ($\mathcal{PT}$ symmetry) or pseudo-chirality into the non-Hermitian Hamiltonian, the number of parameters required for the existence of EP3 can be reduced from four to two.
	To be specific, $\nu$ is set to zero to make the non-Hermitian Hamiltonian satisfy the $\mathcal{PT}$ symmetry.
	The two parameters, $\gamma$ and $h$, ensure that EP3 can exist as isolated points.
	The location of the isolated EP3 is constrained by $h=0$ and $\gamma^2-h^2+2\mu^2=1$.
	Thus, EP3 appears at $h=0, \gamma=\pm\sqrt{1-2\mu^2}$ as shown in Fig.~\ref{Fig1}b (red point).
	Note that this stand-alone EP3 locates at the cusp of two second-order ELs (purple and orange lines in Fig.~\ref{Fig1}b), which is similar to the case experimentally demonstrated recently in the acoustic system\cite{Sci_Tang}.
	In order to observe third-order ELs on the $\gamma-h$ plane, more symmetries are needed to further reduce the number of constraints.
	When both $\mu$ and $\nu$ are set to zero, the non-Hermitian Hamiltonian has both $\mathcal{PT}$ symmetry and pseudo-chirality.
	Then the EP3s appear on the hyperbola satisfying $\gamma^2-h^2=1$.
	As shown in Fig.~\ref{Fig1}c, a third-order EL consisting entirely with EP3s emerges as the red line.

\begin{figure}[http]
			\centering
			\includegraphics[width=1\columnwidth]{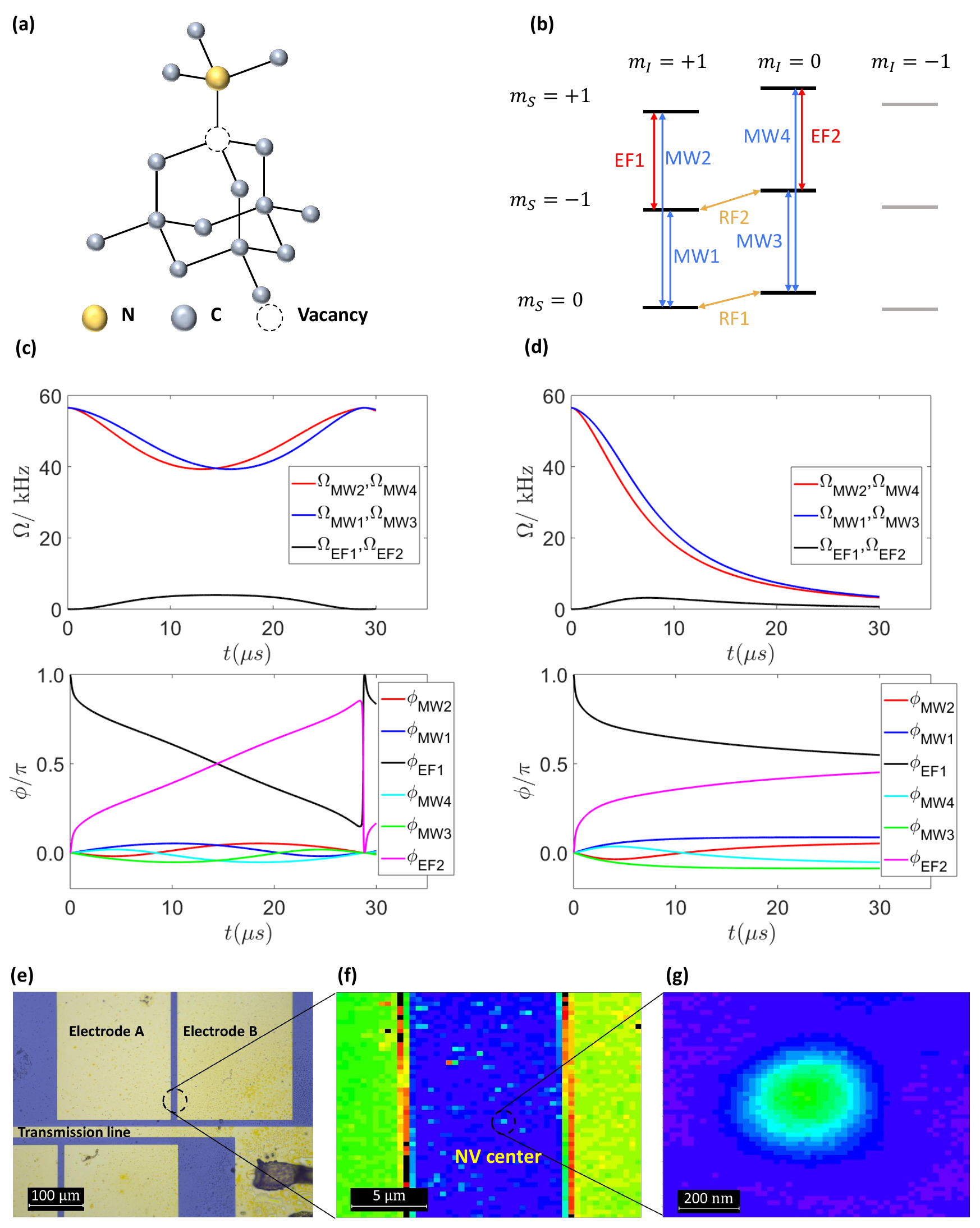}
			\caption{
				Experimental system of an NV center in diamond.
				(a) Atomic structure of the NV center. The gray, yellow and dotted balls stand for the carbon nuclear, the nitrogen nuclear and the vacancy, respectively. (b) The energy levels of the ground state of the NV center with $m_S$ and $m_I$ the electron and nuclear magnetic quantum numbers. The radio-frequency (RF) pulses (orange arrows) are applied to initial state preparation of the nuclear spin. Microwave (MW) (blue arrows) and electric field (EF) pulses (red arrows) are implemented to realize the non-Hermitian Hamiltonian. (c-d) The amplitudes ($\Omega_{\alpha}$) and phases ($\phi_{\alpha}$) of the MW and EF pulses for the parameter configurations $\gamma=0.5$, $h=\mu=\nu=0$ (c) and $\gamma=1$, $h=\mu=\nu=0$ (d). (e-g) Electrodes and transmission line fabricated on the diamond (e), and the location (f) and the photoluminescence image (g) of NV center. The MW and EF pulses are applied by transmission line and electrodes, respectively.
			}
			\label{Fig2}
		\end{figure}

		The NV center, which is an atomic-scale defect in diamond, was utilized to observe the third-order EL (Fig.~\ref{Fig2}a).
		Taking the electron spin in NV center as the system and the nuclear spin as the ancilla qubit, the non-Hermitian Hamiltonian can be realized based on the dilation method \cite{Sci_Wu,PRL_Liu}.
		The subspace spanned by $|m_S\rangle_e|1\rangle_n$ and $|m_S\rangle_e|0\rangle_n$ with $m_S=1, 0, -1$ (Fig.~\ref{Fig2}b) was utilized to construct the dilated Hamiltonian.
		Here subscripts $e$ and $n$ label the electron and nuclear spin states, respectively.
		The dilated Hamiltonian has the form
		\begin{equation}
			H_{\rm tot}(t)=\Gamma(t)\otimes|1\rangle_n~_n\langle1|+\Lambda(t)\otimes|0\rangle_n~_n\langle0|.
		\end{equation}
		Here $\Gamma(t)$ and $\Lambda(t)$ are Hermitian operators on the two subspace of the system,
		\begin{equation}
			\Gamma(t)=\left[\begin{array}{ccc}
				d_1(t) & a_1(t) & c_1(t)\\
				a_1^*(t) & d_2(t) & b_1(t)\\
				c_1^*(t) & b_1^*(t) & d_3(t)
			\end{array}\right],\Lambda(t)=\left[\begin{array}{ccc}
				d_4(t) & a_2(t) & c_2(t)\\
				a_2^*(t) & d_5(t) & b_2(t)\\
				c_2^*(t) & b_2^*(t) & d_6(t)
			\end{array}\right],
		\end{equation}
		where the parameters $a_i(t)$, $b_i(t)$, $c_i(t)~(i=1,2)$ and $d_j(t)~(j=1,...,6)$ are determined by the non-Hermitian Hamiltonian according to the dilation method (See Supplementary Materials, S4 for the derivation and the detailed expressions).
		The state evolution under the non-Hermitian Hamiltonian can be obtained when the evolution of the dilated state under the dilated Hamiltonian is projected into the subspace where the ancilla is $\ket{-}_n=(\ket{0}_n-i\ket{1}_n)/\sqrt{2}$.
		To implement the dilated Hamiltonian, the diagonal elements $d_j(t)$ of $H_{\rm tot}(t)$ were realized by choosing an appropriate interaction picture and the off-diagonal elements $a_1(t)$, $b_1(t)$, $a_2(t)$ and $b_2(t)$ of $H_{\rm tot}(t)$ were realized by four selective microwave (MW) pulses MW2, MW4, MW1 and MW3, respectively (blue arrows in Fig.~\ref{Fig2}b).
		It is noted that there are non-zero off-diagonal elements $c_1(t)$ and $c_2(t)$ in $H_{\rm tot}(t)$ corresponding to the transitions between $|m_S=\pm 1\rangle_e$ which cannot be excited by MW pulses as these transitions are magnetic dipole forbidden.
		To address this issue, alternating current (AC) electric field pulses EF1 and EF2 (red arrows in Fig.~\ref{Fig2}b) were generated by the fabricated electrodes on the diamond surface as shown in Fig.~\ref{Fig2}e to realize the off-diagonal elements $c_1(t)$ and $c_2(t)$, respectively.
		The amplitudes and phases of the MW and electric field pulses were appropriately set according to the off-diagonal elements in $H_{\rm tot}(t)$ (See Supplementary Materials, S4 for the detailed derivation).
		Two examples of the amplitudes and phases of the MW and electric field pulses are shown in Fig.~\ref{Fig2}c and Fig.~\ref{Fig2}d for the parameter configurations $\gamma=0.5$, $h=\mu=\nu=0$ and $\gamma=1$, $h=\mu=\nu=0$, respectively.
		The evolution under the non-Hermitian Hamiltonian is characterized by the normalized population of the electron spin state $|0\rangle_e|1\rangle_n$ (denoted as $P_0$), which was obtained by $P_0=P_{|0\rangle_e|1\rangle_n}/\sum_{m_S=1,0,-1}P_{|m_S\rangle_e|1\rangle_n}$, with $P_{|m_S\rangle_e|1\rangle_n}$ being the populations of state $|m_S\rangle_e|1\rangle_n$ $(m_S=1,0,-1)$.
		
		The target non-Hermitian Hamiltonians were realized with precisely controlled parameters $\gamma$, $h$, $\mu$ and $\nu$.
		The parameters $\mu$ and $\nu$ that control the symmetry of the Hamiltonian can be retrieved independently by measuring the conserved quantities related to symmetries.
		Taking the case of $\mu=\nu=0$ as an example, the parameters can be experimentally obtained as $\mu_{\rm exp}=0.01(2)$ and $\nu_{\rm exp}=0.01(2)$ through the specific procedure described in Methods, as shown in Extended Data Fig.~\ref{Extend_Fig1}(a-b).
		This example and the data for other $\mu$ and $\nu$ in the Supplementary Materials, S6 show that the experimental results are consistent with the theoretical predictions, proving that the parameters $\mu$ and $\nu$ can be precisely controlled.
		The parameters $\gamma$ and $h$ are retrieved experimentally from the population evolutions of the non-Hermitian Hamiltonian started with two different initial states and measured at different measurement bases.
		We take the case $h=0$, $\gamma=1$ (when $\mu=\nu=0$) as an example.
		The two sets of initial states and the measurement bases are chosen as $\ket{\psi_1} = (0,1,0)^T$, $\ket{\phi_1} = (0,1,0)^T$ and $\ket{\psi_2} = (-i/\sqrt{2},1/\sqrt{2},0)^T$, $\ket{\phi_2} = (1/\sqrt{2},1/\sqrt{2},0)^T$.
		Fitting simultaneously the two population evolutions in Extended Data Fig.~\ref{Extend_Fig1}(c-d) leads to $h_{\rm exp}=0.00(3)$, $\gamma_{\rm exp}=1.00(3)$, which are in good agreement with the corresponding assigned parameters.
		
\begin{figure}[http]
			\centering
			\includegraphics[width=1\columnwidth]{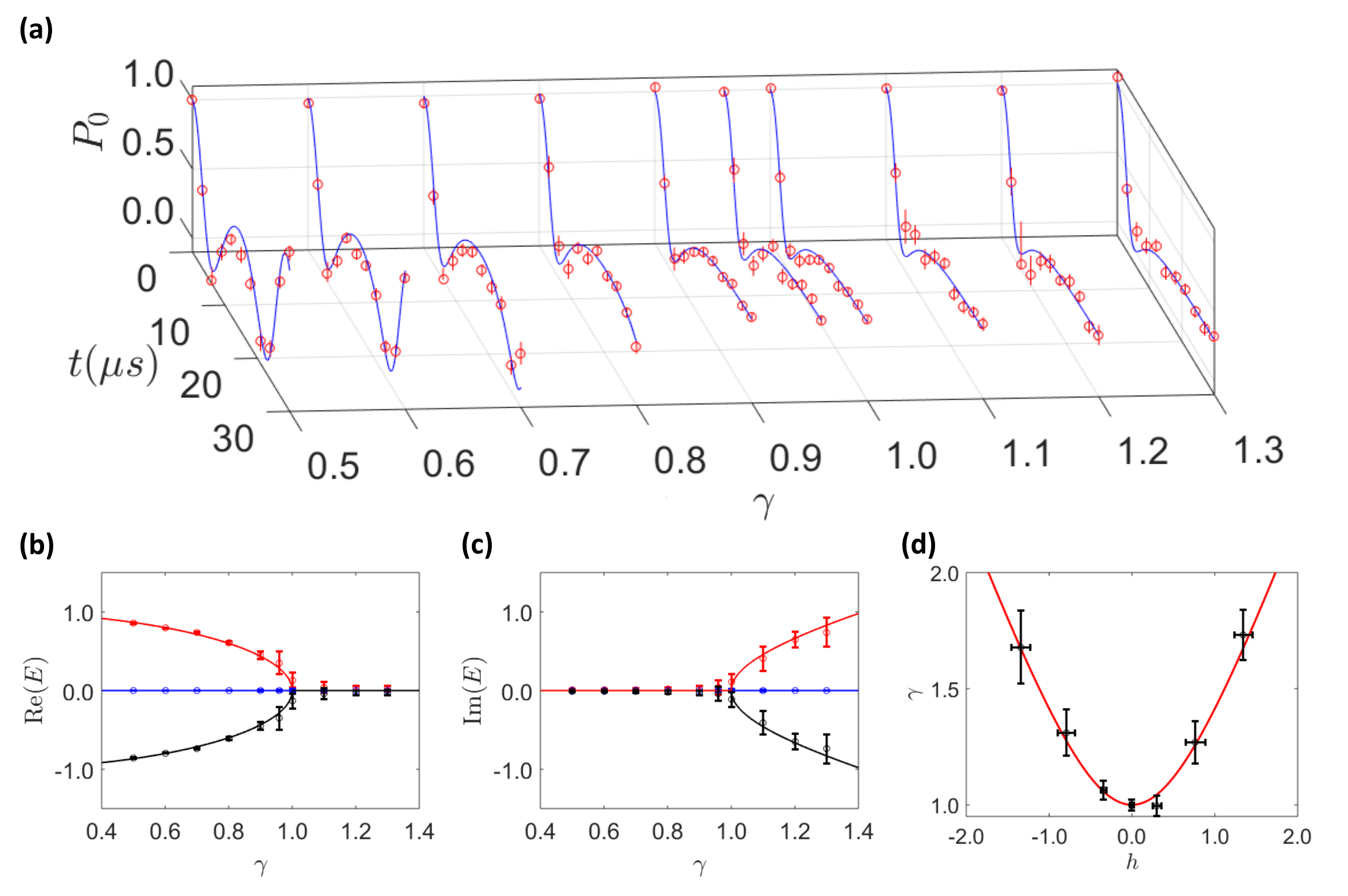}
			\caption{Observation of the third-order exceptional line. (a) Experimental state evolution under non-Hermitian Hamiltonian when $h=\mu=\nu=0$ with different values of parameter $\gamma$. $P_0$ is the renormalized population of the state $|0\rangle_e$ of the electron spin when the nuclear spin state is $|1\rangle_n$. Red dots with error bars are experimental results, and blue lines are the theoretical predictions. (b-c) The real (b) and imaginary (c) parts of the three eigenvalues of the non-Hermitian Hamiltonian.
				The red, blue and black dots with error bars are the three eigenvalues calculated from the non-Hermitian Hamiltonian with parameters $\gamma_{\rm exp}$, $h_{\rm exp}$, $\mu_{\rm exp}$ and $\nu_{\rm exp}$, respectively. (d) The observation of the third-order exceptional line. The red line is the theoretical prediction of third-order EL, and black dots with error bars stand for the experimentally observed EP3s. }
		\label{Fig3}
	\end{figure}

		To observe the third-order EL, we experimentally measured the evolution of the state under the non-Hermitian Hamiltonian and verified the degeneracies of the eigenvalues and eigenstates at the EP3s.
		Figure~\ref{Fig3}a shows the state evolution under non-Hermitian Hamiltonians which satisfy both $\mathcal{PT}$ symmetry and pseudo-chirality with various values of $\gamma$ when $h =\mu=\nu=0$.
		The state evolution under the non-Hermitian Hamiltonian is demonstrated by recording $P_0$ in the time varied from 0 to 30 $\mathrm{\mu s}$.
		All errors shown are one standard deviation with one million averages.
		The measured dependence of the state evolution on the parameter $\gamma$ shows good agreement with the theoretical prediction.
		When $|\gamma|<1$, the Hamiltonian with real eigenvalues determines the oscillatory dynamics.
		When $|\gamma|>1$, the imaginary eigenvalues break down the oscillation and make the evolution approach a steady state.
		To observe the EP3, the eigenvalues are calculated from the non-Hermitian Hamiltonian with parameters $\gamma_{\rm exp}$, $h_{\rm exp}$, $\mu_{\rm exp}$ and $\nu_{\rm exp}$.
		Figure~\ref{Fig3}b and \ref{Fig3}c shows the eigenvalues of the non-Hermitian Hamiltonian and the experimental results agree well with the theoretical predictions.
		The eigenvalue $E_2$ is always zero with different $\gamma$.
		The other two eigenvalues $E_1$ and $E_3$ are real when $\gamma<1$.
		When $\gamma>1$, the real parts of the two eigenvalues $E_1$ and $E_3$ vanish and the imaginary parts appear.
		The EP3 was observed at $\gamma=1$, where all the three eigenvalues coalesce to zero.
		Furthermore, the EP3 is verified by the coalescence of the three eigenstates which are obtained by a method similar to that in the literature\cite{npj_Yu} (See Supplementary Materials, S7).
		The overlap between two of the eigenstates is characterized by the fidelity, $F_{ij}$, between the eigenstates corresponding to the eigenvalues $E_i$ and $E_j$.
		At the EP3, the experimental results of $F_{ij}$ are $F_{12}=0.99(2)$, $F_{13}=0.99(2)$ and $F_{23}=1.00(2)$, which demonstrates the degeneracy of the three eigenstates.
		The triple degeneracy of the eigenstates and the corresponding eigenvalues identify the EP3.
		For the observation of third-order EL, we set different values of $\gamma$ and $h$ which satisfy the relation $\gamma^2-h^2=1$.
		The observed EP3s are identified by the coalescence of the three eigenvalues.
		Figure~\ref{Fig3}d shows the experimental results of EP3s together with the theoretical predictions.
		The experimentally identified EP3s (black dots with error bars) agree with the theoretical predicted third-order EL (red line) in the $h-\gamma$ plane.
		All the three mutual overlaps between the eigenstates equal one within the margin of error as listed in Table~\ref{Fidelity}, showing the degeneracy of the eigenstates of EP3s on the third-order EL.
		Thus, the third-order EL has been experimentally demonstrated.
		
\begin{figure}[http]
			\centering
			\includegraphics[width=1\columnwidth]{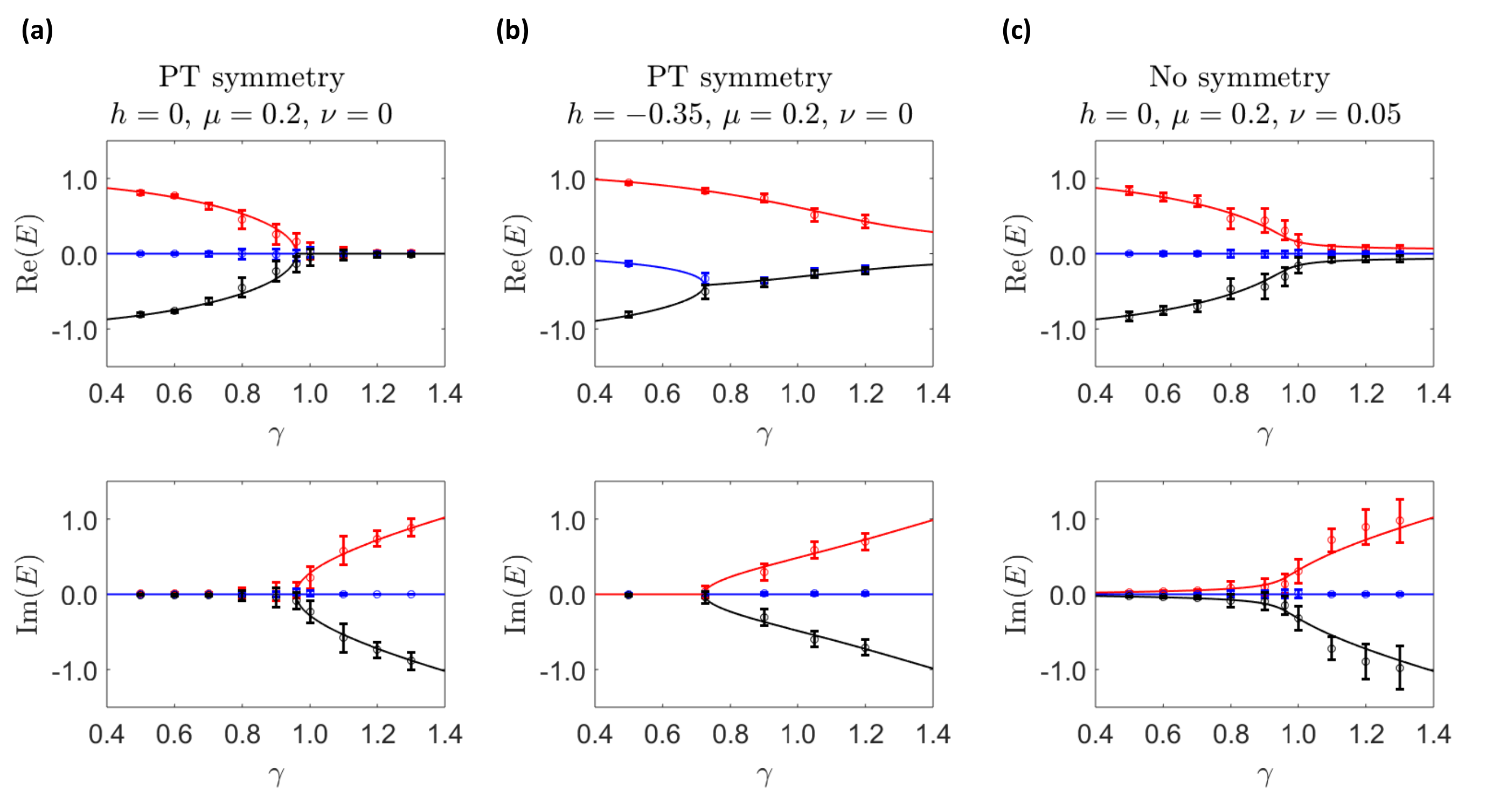}
			\caption{Behavior of the EP3 in the two-dimensional parameter space with different symmetries. (a) When only $\mathcal{PT}$ symmetry is preserved, EP3 exists with $h=0$. (b) EP3 disappears with none-zero value of $h$. (c) When both symmetries are broken, EP3 does not exist even when $h=0$.
				The red, blue and black dots with error bars are the three eigenvalues calculated from the non-Hermitian Hamiltonian with parameters $\gamma_{\rm exp}$, $h_{\rm exp}$, $\mu_{\rm exp}$ and $\nu_{\rm exp}$. The lines of the corresponding colors are theoretical predictions.}
			\label{Fig4}
		\end{figure}

		Next, we show the relationship between the behavior of EP3 and the symmetries of the non-Hermitian Hamiltonians.
		In Fig.~\ref{Fig4}a and \ref{Fig4}b, the pseudo-chirality of the system is broken with the perturbation $\mu=0.2$, while the $\mathcal{PT}$ symmetry is still preserved with $\nu=0$.
		The measurements of the conserved quantities give the results of $\mu_{\rm exp}=0.20(3)$ and $\nu_{\rm exp}=0.01(2)$ which are consist with the preset values.
		The behavior of the eigenvalues as a function of parameter $\gamma$ is similar to the case with both the two symmetries unbroken when $h=0$ (Fig.~\ref{Fig4}a).
		The EP3 can be identified at $\gamma=0.96$, which satisfies the constraint $\gamma^2+2\mu^2=1$.
		The degeneracy of the eigenstates at the EP3 is demonstrated by the fidelities in Table~\ref{Fidelity}.
		When $h=-0.35$, there is no EP3 (Fig.~\ref{Fig4}b).
		However, an EP2 can be observed with $\gamma=0.73$, where only two eigenvalues $E_2$ and $E_3$ coalesce to the same value and the corresponding eigenstates degenerate (Table~\ref{Fidelity}).
		This result shows that when only $\mathcal{PT}$ symmetry is held in the non-Hermitian Hamiltonian, the third-order EP structure is reduced from lines to isolated points.
		We further introduced a perturbation $\nu=0.05$ to break the $\mathcal{PT}$ symmetry of the non-Hermitian Hamiltonian, which is verified by the experimental retrieved $\nu_{\rm exp}=0.05(1)$.
		Experimental results show that the three eigenvalues will not coalesce by changing the value of $\gamma$ (Fig.~\ref{Fig4}c).
		The apparent deviations of the overlaps between the eigenstates from one verify that there is no degeneracy of the eigenstates (Table~\ref{Fidelity}).
		Therefore, no EP3 has been observed when the non-Hermitian Hamiltonian has no symmetry.
		
	\begin{table*}[http]\centering
		\caption{\textbf{Experimental overlaps between the eigenstates.} The fidelity $F_{ij}$ that characterizes the overlap between two of the eigenstates corresponding to the eigenvalues $E_i$ and $E_j$. The values of $F_{ij}$ shown in bold fonts are expected to equal one.}
		\textrm{\\}
		\renewcommand{\multirowsetup}{\centering}
		\begin{threeparttable}

			\begin{tabular}{c|c |ccc}  
				\hline\hline
				Symmetry and parameters $\mu$ and $\nu$ & Parameters $h$ and $\gamma$ & $F_{12}$ & $F_{13}$ & $F_{23}$ \\ \hline  
				\multirow{7}{*}{\tabincell{c}{PT symmetry \& pesudo-chirality \\ $\mu=0$, $\nu=0$}} & $h=0$, $\gamma=1$ & \bf 0.99(2) & \bf 0.99(2) & \bf 1.00(2) \\
				& $h=0.35$, $\gamma=1.06$ & \bf 0.99(3) & \bf 0.98(2) & \bf 0.98(2)\\
				& $h=-0.35$, $\gamma=1.06$ & \bf 1.00(3) & \bf 0.98(3) & \bf 0.98(2)\\
				& $h=0.75$, $\gamma=1.25$ & \bf 0.98(2) & \bf 0.98(2) & \bf 0.98(2)\\
				& $h=-0.75$, $\gamma=1.25$ & \bf 0.99(2) & \bf 0.99(2) & \bf 0.98(2)\\
				& $h=1.4$, $\gamma=1.72$ & \bf 0.98(2) & \bf 0.98(2) & \bf 0.98(2)\\
				& $h=-1.4$, $\gamma=1.72$ & \bf 0.98(2) & \bf 0.98(2) & \bf 0.98(2)\\  \hline
				
				\multirow{2}{*}{\tabincell{c}{PT symmetry \\ $\mu=0.2$, $\nu=0$}} & $h=0$, $\gamma=0.96$ & \bf 0.98(2) & \bf 0.98(2) & \bf 0.98(2) \\
				& $h=-0.35$, $\gamma=0.73$ & 0.26(5) & 0.21(6) & \bf 0.98(3)\tnote{[1]}\\ \hline
				
				\tabincell{c}{No symmetry \\ $\mu=0.2$, $\nu=0.05$} & $h=0$, $\gamma=0.96$ & 0.91(4) & 0.72(5) & 0.92(4)\tnote{[2]} \\ \hline\hline
				
			\end{tabular}
			\begin{tablenotes}
				\footnotesize
				\item[{[1]}] The theoretical predictions of the fidelities for the case $h=-0.35$, $\gamma=0.73$ when PT symmetry is preserved are $F_{12}=0.22$, $F_{13}=0.22$ and $F_{23}=1$.
				\item[{[2]}] The theoretical predictions of the fidelities for the case $h=0$, $\gamma=0.96$ without symmetry are $F_{12}=0.94$, $F_{13}=0.77$ and $F_{23}=0.94$.
			\end{tablenotes}
			
			\label{Fidelity}
		\end{threeparttable}
	\end{table*}
		
		In conclusion, we experimentally investigated the third-order EP structure with different symmetries at the atomic scale.
		With both $\mathcal{PT}$ symmetry and pseudo-chirality introduced, the third-order EL has been successfully observed.
		If the pseudo-chirality is broken, the third-order EL will reduce to isolated EP3s.
		When $\mathcal{PT}$ symmetry is further removed, the EP3 disappears.
		Our results indicate that symmetries play an important role in the high-order EP structure.
		As noted by recent theoretical work, the presence of certain symmetries decreases the codimension of EPs \cite{PRL_Deplace}, which means that fewer tuning parameters are required to construct a high-order EP.
		The existence of the third-order EL not only makes the high-order EP related phenomena more robust \cite{PRL_QZhong,Laser_Photonics_Rev,NC_SS}, but also introduces anisotropy in the parameter space \cite{PRL_Tang_W,PRL_Ding_K} in the sense that the splitting of eigenvalues and eigenstates occurs only in parameter variations perpendicular to the EP lines.
		As shown in Ref.~\onlinecite{Nat_Patil}, the complicated EP2 knots can introduce intriguing winding phenomena, and here similar structures for high-order EPs may be constructed under certain symmetries.
		Similar topological characterization of the third-order EL in our case can be realized by the difference between the sign of the resultants for two points in the two-dimensional parameter space following a method similar to that in literatures\cite{PRB_topo,PRB_topo2} (See Supplementary Materials, S8 for detailed discussion).
		In addition to showing that the NV center is a desirable atomic-scale platform for investigation non-Hermitian quantum physics, our work not only facilitates the investigation of the topological properties of high-order EPs\cite{PRL_Deplace,PRL_Mandal} but also provides guidance for novel quantum control methods and enhanced quantum sensing utilizing high-order EPs \cite{NC_Lau,PR_Wiersig}.
		While preparing
		this manuscript, we became aware of a related work in an
		acoustic system \cite{NC_Tang_W}.
	
	\section{methods}
		
		\subsection{Experiment setup}
		
		\quad \ \ The experiments were implemented on a [100] oriented NV center in an isotopically purified ([$^{12}$C]=99.999\%) diamond synthesized by the chemical vapor deposition method.
		The Hamiltonian of the NV center can be written as $H_{\mathrm{NV}} = 2\pi(DS_z^2 + \omega_eS_z + QI_z^2 + \omega_nI_z + AS_zI_z)$,
		where $D=2.87$ GHz is the zero-field splitting of the electron spin, $Q=-4.95$ MHz is the nuclear quadrupolar interaction, and $A=-2.16$ MHz is the hyperfine coupling between the electron spin and the nuclear spin.
		$\omega_e$ ($\omega_n$) denotes the Zeeman splitting of the electron (nuclear) spin.
		$S_z$ and $I_z$ are the spin-1 operators of the electron spin and the nuclear spin, respectively.
		The dephasing time of the electron spin of the NV center in our experiments was measured to be $T_2^\star = 127(4)$ $\mathrm{\mu s}$, which is long enough comparing to the evolution time
		of the order of 30 $\mathrm{\mu s}$ (See Supplementary Materials, S4).
		The diamond was mounted on an optically detected magnetic resonance setup.
		The optical excitation was realized by the 532 nm green laser pulses modulated by an acousto-optic modulator (ISOMET).
		The laser beam traveled twice through the acousto-optic modulator before going through an oil objective (Olympus, UPLXAPO 100*O, NA 1.45).
		The phonon sideband fluorescence (wavelength, 650-800nm) from the NV center went through the same oil objective and was collected by an avalanche photodiode (Perkin Elmer, SPCM-AQRH-14) with a counter card.
		The magnetic field of 500 G was provided by a permanent magnet along the NV symmetry axis.
		An arbitrary waveform generator (Keysight M8190A) generated MW, electric field and radio-frequency (RF) pulses to manipulate the states of the NV center.
		The MW, electric field and RF pulses were amplified individually by power amplifiers (two Mini Circuits ZHL-15W-422-S+ for MW and electric field pulses and LZY-22+ for RF pulses).
		The MW pulses and electric fields were applied by transmission line and electrodes fabricated on the surface of diamond, respectively.
		The RF pulses were carried by a home-built RF coil.
		
		\begin{figure}[http]
			\centering
			\includegraphics[width=1\columnwidth]{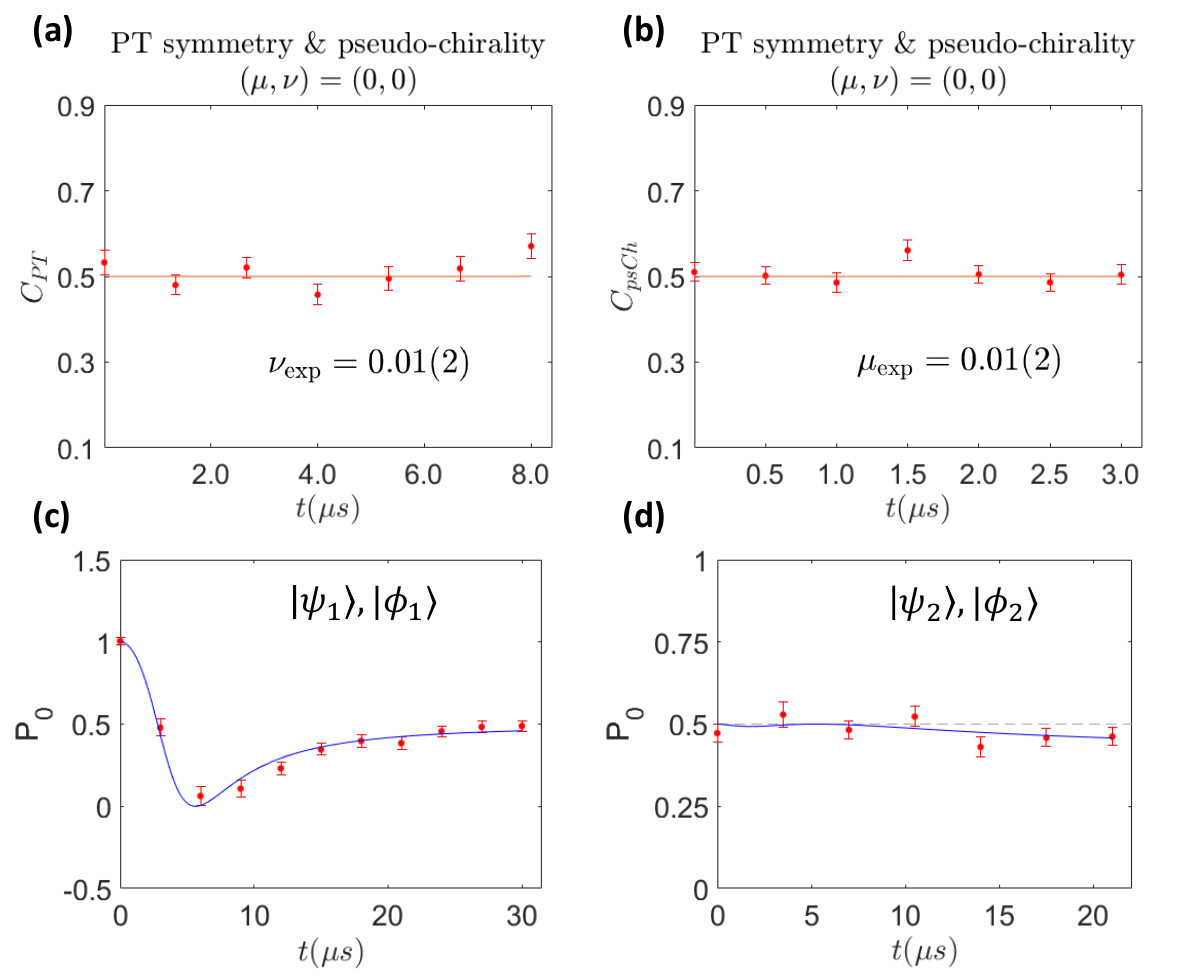}
			\caption{Experimental measurement of the parameters $\gamma_{\rm exp}$, $h_{\rm exp}$, $\mu_{\rm exp}$ and $\nu_{\rm exp}$ of the non-Hermitian Hamiltonian. The case of $\mu=0$, $\nu=0$, $h=0$ and $\gamma=1$ is taken as an example. (a) The measured quantities $C_{\mathcal{PT}}$ and the retrieved parameter $\nu_{\rm exp}$. (b) The measured quantities $C_{\rm psCh}$ and the retrieved parameter $\mu_{\rm exp}$. (c-d) Population evolution under $H^{(\mu,\nu)}(\gamma,h)$ with two different sets of initial states and measurement bases. $\mu_{\mathrm{exp}}$ and $\nu_{\mathrm{exp}}$ are obtained via linear fitting, and $\gamma_{\rm exp}$ and $h_{\rm exp}$ are obtained from the evolution as described in the text. Red dots with error bars are experimental results, and blue lines are the theoretical predictions.}
			\label{Extend_Fig1}
		\end{figure}
		
		\subsection{Acquisition of the eigenvalues of the non-Hermitian Hamiltonian}
		\quad \ \  The complex eigenvalues and non-orthogonal eigenstates of the non-Hermitian Hamiltonians in study make it challenging to directly determine the eigenvalues without relying on a model.
		Here we calculated the eigenvalues from the experimentally retrieved parameters $\gamma_{\rm exp}$, $h_{\rm exp}$, $\mu_{\rm exp}$ and $\nu_{\rm exp}$ based on the form of non-Hermitian Hamiltonian $H^{(\mu,\nu)}(\gamma,h)$ as shown in Eq.~\ref{Hamiltonian}.
		First, the parameter $\nu$ is obtained by measuring the $\mathcal{PT}$ symmetry conserved quantity defined as $C_{\mathcal{PT}}(t)=|\bra{\phi}e^{isH^*t}U_{\mathcal{PT}}e^{-isHt}\ket{\psi}|^2$, where $U_{\mathcal{PT}}$ is such a unitary operator that the Hamiltonian $H_{PT}$ with $\mathcal{PT}$ symmetry satisfies $U_{\mathcal{PT}}H_{\mathcal{PT}}U_{\mathcal{PT}}^{-1}=H_{\mathcal{PT}}^*$ and $s$ is a nonzero coefficient.
		By choosing $\ket{\psi} = (0,(1+i)/2,1/\sqrt{2})^T$ and $\ket{\phi} = ((1+i)/2,1/\sqrt{2},0)^T$, we obtain $C_{\mathcal{PT}}(t) = 1/2+s\nu t+O(t^2)$.
		In our experiment, $s$ is set as $2\pi\times30$ kHz.
		The parameter $\nu$ can be obtained from $C_{\mathcal{PT}}(t)$, which is independent of $\mu$, $\gamma$ and $h$ to the first order of $t$.
		Here, we take the case of $\nu=0$ as an example.
		The experimental result $\nu_{\rm exp}=0.01(2)$ as shown in Extended Data Fig.~\ref{Extend_Fig1}a is in good agreement with the theoretical prediction.
		The cases of nonzero $\nu$ can be found in Supplementary Materials, S6.
		Second, the parameter $\mu$ is retrieved from conserved quantity related to pseudo-chirality defined as $C_{\rm psCh}(t)=|\bra{\phi}e^{-isH^\dagger t}U_{\rm psCh}e^{-isHt}\ket{\psi}|^2$, where $U_{\rm psCh}$ makes the Hamiltonian satisfy the condition $U_{\rm psCh}H_{\rm psCh}U_{\rm psCh}^{-1}=-H_{\rm psCh}^{\dag}$ in presence of pseudo-chirality.
		The choice of $\ket{\psi} = (0,1,0)^T$ and $\ket{\phi} = (0,1/\sqrt{2},i/\sqrt{2})^T$ leads to $C_{\mathcal{\rm psCh}}(t) = 1/2+2s\mu t+O(t^2)$.
		In measuring $C_{\rm psCh}\left(t\right)$, $s$ is set as $2\pi\times20$ kHz.
		We take $\mu=0$ as an example.
		As shown in the Extended Data Fig.~\ref{Extend_Fig1}b, the experimental $\mu$ value of $0.01\left(2\right)$ is in good agreement with the target value (See Supplementary Materials, S6 for more relevant data).
		Third, the parameters $\gamma$ and $h$ are determined experimentally from the population evolutions of the non-Hermitian Hamiltonian started with two different initial states and measured at different measurement bases.
		The population evolution is characterized by $P_0(t) = |\bra{\phi}e^{-iHt}\ket{\psi}|^2/N(t)$, where $N(t) = |\bra{\psi}e^{iH^\dagger t}e^{-iHt}\ket{\psi}|^2$ is a normalization factor.
		The two sets of initial states and the measurement bases are chosen as $\ket{\psi_1} = (0,1,0)^T$, $\ket{\phi_1} = (0,1,0)^T$ and $\ket{\psi_2} = (-i/\sqrt{2},1/\sqrt{2},0)^T$, $\ket{\phi_2} = (1/\sqrt{2},1/\sqrt{2},0)^T$.
		We take the case $h=0,\gamma=1$ when $\mu=\nu=0$ as an example.
		Fitting simultaneously the population evolutions to theoretical predictions under the non-Hermitian Hamiltonian leads to $h_{\rm exp}=0.00\left(3\right)$ and $\gamma_{\rm exp}=1.00\left(3\right)$, which are in good agreement with the corresponding assigned parameters (Extended Data Fig.~\ref{Extend_Fig1}c-d) .
		The uncertainties of $h_{\rm exp}$ and $\gamma_{\rm exp}$ come from the statistical error of the experiment results (See Supplementary Materials, S6 for more relevant data).
		The eigenvalues are evaluated with $(\mu_{\rm exp},\nu_{\rm exp},h_{\rm exp},\gamma_{\rm exp})$ using the model Hamiltonian in Eq.\ref{Hamiltonian}.
		The errorbars of the eigenvalues are obtained from $(\mu_{\rm exp},\nu_{\rm exp},h_{\rm exp},\gamma_{\rm exp})$ and their standard deviations based on the Monte Carlo method.
	
	This work was supported by the National Key R\&D Program of China (Grant No. 2018YFA0306600), the National Natural Science Foundation of China (Grant Nos. 12174373, 92265204 and 12261160569), the Chinese Academy of Sciences (Grant Nos. XDC07000000 and GJJSTD20200001), the Innovation Program for Quantum Science and Technology (Grant No. 2021ZD0302200), the Anhui Initiative in Quantum Information Technologies (Grant No. AHY050000) and the Hefei Comprehensive National Science Center. Ya Wang thanks the Fundamental Research Funds for the Central Universities for their support. W. L. is funded by Beijing University of Posts and Telecommunications Innovation Group. This work was partially carried out at the USTC Center for Micro and Nanoscale Research and Fabrication.

	\onecolumngrid
	\vspace{1.5cm}
	\begin{center}
		\textbf{\large Supplementary Material}
	\end{center}
	
	\setcounter{figure}{0}
	\setcounter{equation}{0}
	\setcounter{table}{0}
	\makeatletter
	\renewcommand{\thefigure}{S\arabic{figure}}
	\renewcommand{\theequation}{S\arabic{equation}}
	\renewcommand{\thetable}{S\arabic{table}}
	\renewcommand{\bibnumfmt}[1]{[RefS#1]}
	\renewcommand{\citenumfont}[1]{RefS#1}
	
	

	\section{S1. Symmetries and the third-order EP structures}
	In this section we provide the details of the third-order EP structures and discuss the role played by symmetries.
	The occurrence of EP3 can be formulated using the language of resultants \cite{Resultant}.
	For a general non-Hermitian (NH) Hamiltonian, the resultants are defined as
	\begin{equation}
		R_1=\mathcal{R}_{p,p'}= \text{det}(S_{p,p'}),R_2=\mathcal{R}_{p',p''}= \text{det}(S_{p',p''}),
	\end{equation}
	where $p(\lambda)=\text{det}[\lambda I-H]$ is the characteristic polynomial of the Hamiltonian, $p'$ $(p'')$ is the first (second) derivative of $p$ and $S_{r,s}$ is the Sylvester matrix between two polynomials $r$ and $s$. For example, if $r(x) = \sum_{i=0}^{3}r_ix^i,s(x) = \sum_{i=0}^{2}s_ix^i$, then
	\begin{equation}
		S_{r,s} = \left[\begin{array}{ccccc}
			r_3 & r_2 & r_1 & r_0 & 0\\
			0 & r_3 & r_2 & r_1 & r_0\\
			s_2 & s_1 & s_0 & 0 & 0\\
			0 & s_2 & s_1 & s_0 & 0\\
			0 & 0 & s_2 & s_1 & s_0
		\end{array}\right].
	\end{equation}
	The conditions for the occurrence of EP3 are
	\begin{equation}
		\text{Re}(R_1)=\text{Im}(R_1)=\text{Re}(R_2)=\text{Im}(R_2)=0.
		\label{Cond}
	\end{equation}
	
	Herein, we take 3-level systems as an example to show how symmetries influence the codimension of EP3. The characteristic polynomial of a 3$\times$3 Hamiltonian takes the form
	\begin{equation}
		p(\lambda) = (\lambda-E_1)(\lambda-E_2)(\lambda-E_3) = f_3\lambda^3+f_2\lambda^2+f_1\lambda+f_0,
	\end{equation}
	where $E_i$ are the eigenvalues of the Hamiltonian. If no symmetry is specified, all the $E_i$, $f_i$ and thus $R_{1,2}$ are complex functions of the parameters. Then Eq. \ref{Cond} gives four constraints for the occurrence of EP3. So the EP3 has codimension 4 in absence of symmetry. If the system has $\mathcal{PT}$ symmetry, the Hamiltonian satisfies $U_{\mathcal{PT}}HU_{\mathcal{PT}}^{-1}=H^*$ for a unitary operator $U_{\mathcal{PT}}$. Then $\text{det}[\lambda I-H] = \text{det}[U_{\mathcal{PT}}(\lambda I-H)U_{\mathcal{PT}}^{-1}]$, and thus
	\begin{equation}
		p(\lambda) = (\lambda-E_1)(\lambda-E_2)(\lambda-E_3) = (\lambda-E_1^*)(\lambda-E_2^*)(\lambda-E_3^*).
	\end{equation}
	Hence the $f_i$ must be real. In this case, the imaginary parts of $R_{1,2}$ vanish, and thus only the two constraints corresponding to the real parts of $R_{1,2}$ are non-trivial. This gives codimension 2 for the EP3 under $\mathcal{PT}$ symmetry. If pseudo-chirality is satisfied, the Hamiltonian satisfies $U_{\rm psCh}HU_{\rm psCh}^{-1}=-H^\dagger$ for a unitary operator $U_{\rm psCh}$. Then
	\begin{equation}
		p(\lambda) = (\lambda-E_1)(\lambda-E_2)(\lambda-E_3) =(\lambda+E_1^*)(\lambda+E_2^*)(\lambda+E_3^*).
	\end{equation}
	Hence $f_1,f_3$ are real and $f_0,f_2$ are purely imaginary. If we consider the Hamiltonian $iH$, then by $U_{\rm psCh}(iH)U_{\rm psCh}^{-1}=(iH)^\dagger$, the resultants for $iH$ are real, which recovers the $\mathcal{PT}$ symmetric case. Since $H$ and $iH$ have exactly the same conditions for the formation of EP3, we obtain that the codimension of EP3 is still 2. When the system has both symmetries, the coefficients $f_0$ and $f_2$ have to be both real and imaginary, i.e., they must vanish. Hence we obtain $p(\lambda) = \lambda^3 + f_1\lambda$ with $f_1$ real ($f_3 = 1$ by definition). Then the only constraint for occurrence of EP3 is simply $f_1=0$. And the codimension is 1 in presence of both symmetries.
	
	For our model, the NH Hamiltonian has the form
	\begin{equation}
		H^{(\mu,\nu)}(\gamma,h)=S_x+(i\gamma+\nu) S_z+hS_y+\mu \left[\begin{array}{ccc}
			0 & 1 & 0\\
			-1 & 0 & -1\\
			0 & 1 & 0
		\end{array}\right].
		\label{model}
	\end{equation}
	Here, $S_{x,y,z}$ are the spin-1 operators and the $\mu,\nu$ terms describe the perturbation.
	The characteristic polynomial for $H^{(\mu,\nu)}(\gamma,h)$ has the form $p(\lambda) = \lambda^3+f_1\lambda+f_0$, where
	\begin{equation}
		f_1 = \gamma^2-1-h^2-2i\gamma \nu-\nu^2+2\mu^2, f_0 = 2\sqrt{2}\mu h(\gamma-i\nu).
	\end{equation}
	And the explicit expressions of $R_{1,2}$ are
	\begin{equation}
		R_1=4f_1^3+27f_0^2,R_2=36f_0.
	\end{equation}
	
	In absence of perturbation, i.e, $\mu=\nu=0$, the model satisfies both  $\mathcal{PT}$ symmetry and pseudo-chirality,
	with the unitary operators chosen as
	\begin{equation}
		U_{\mathcal{PT}} = \left[\begin{array}{ccc}
			0 & 0 & 1\\
			0 & 1 & 0\\
			1 & 0 & 0
		\end{array}\right], U_{\rm psCh} = \left[\begin{array}{ccc}
			1 & 0 & 0\\
			0 & -1 & 0\\
			0 & 0 & 1
		\end{array}\right].
		\label{Unitaries}
	\end{equation}
	In this case only the constraint $\text{Re}(R_1(\gamma,h))=0$ needs to be considered. Since $f_1 = \gamma^2-1-h^2$ and $f_0 = 0$, $R_1(\gamma,h)$ reduces to $4(\gamma^2-1-h^2)^3$. Hence the constraint is equivalent to $\gamma^2-1-h^2=0$. Thus the EP3s form a curve in the $h$-$\gamma$ plane. When $\mu\ne0$, the pseudo-chirality is broken and the $\mathcal{PT}$ symmetry is preserved. As noted above, the $\mathcal{PT}$ symmetry renders both $R_1$ and $R_2$ to be real, and thus the conditions for EP3 can be simplified to
	\begin{equation}
		\gamma h=0,\gamma^2-1-h^2+2\mu^2=0.
	\end{equation}
	We note that only the $\gamma\ne0$ case is considered (or the original Hamiltonian would be a trivial Hermitian one), and thus the EP3 must be the intersection of the $h=0$ axis and the $\gamma^2-1-h^2+2\mu^2=0$ curve in the $h$-$\gamma$ plane. If both $\mu$ and $\nu$ are non-zero, both symmetries are broken, and then the codimension of EP3 is greater than the dimension of the parameter space. Thus, in this case the parameters $\gamma$ and $h$ can not fulfill (non-trivially) the four following conditions for non-zero values of $\mu$ and $\nu$:
	\begin{equation}
		\gamma^2-1-h^2-\nu^2+2\mu^2 = \gamma\nu = \gamma\mu h = \nu\mu h = 0.
	\end{equation}
	From the equation $R_2 = 36f_0 = 0$, the EP3 (if any) can only locate in the $h=0$ axis. Thus we focus on the cases with $h = 0$ in Fig.4b in the main text.
	\section{S2. The dispersion relation near the EL3}
\begin{figure}
		\centering
		
		\includegraphics[width=1.0\columnwidth]{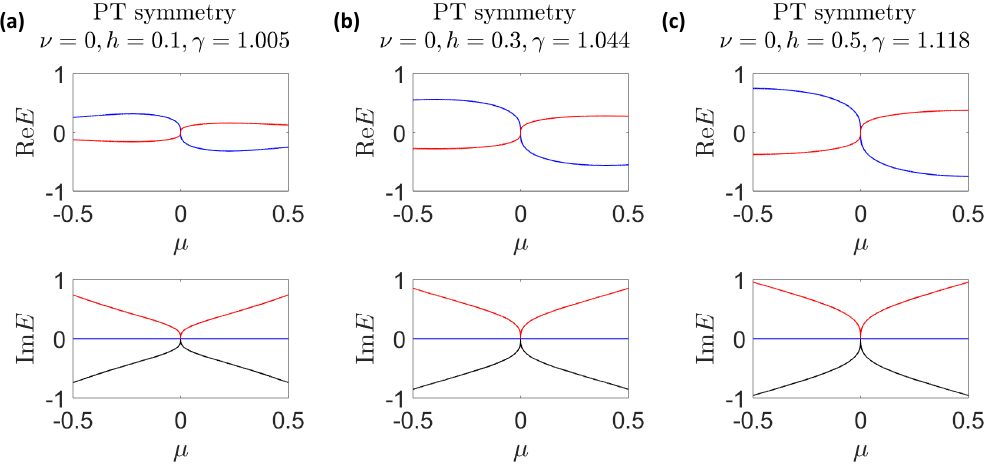}
		\caption{The change of the real and imaginary parts of the three eigenvalues with parameter $\mu$ for three NH Hamiltonians having PT symmetry.
		}
		\label{Cubic dispersion}
	\end{figure}

	The EP3 considered in our model can exhibit both square-root and cube-root dispersion relations with respect to different parameters. To demonstrate the cube-root dispersion, we focus on the dispersion with respect to $\mu$ near the EL3 in the main text. In this case $\gamma,h,\mu\ne0$ and $\nu = 0$. Thus the characteristic polynomial has the form
	\begin{equation}
		p(\lambda)=\lambda^3+(\gamma^2-1-h^2+2\mu^2)\lambda+2\sqrt{2}\mu h\gamma.
	\end{equation}
	At the EL3, we consider $(h,\gamma) = (h_0,\gamma_0)$ such that $\gamma_0^2 = 1+h_0^2$. Then the characteristic polynomial is reduced to $\lambda^3+2\mu^2\lambda+2\sqrt{2}\mu h_0\gamma_0$. The discriminant is $\Delta = 8\mu^6/27+2h_0^2\mu^2\gamma_0^2$, which is positive. Thus by the Cardano's formula, we have
	\begin{equation}
		E_i = \omega^i (\epsilon_+)^{1/3}+\omega^{2i} (\epsilon_-)^{1/3}, i=1,2,3,
	\end{equation}
	where $\omega = e^{2i\pi/3}$ is the unit cube root and
	\begin{equation}
		\epsilon_\pm = -\sqrt{2}h_0\mu\gamma_0\pm\sqrt{2}|h_0\mu\gamma_0|\sqrt{1+4\mu^4/27h_0^2\gamma_0^2}.
	\end{equation}
	Without loss of generality, assume that $\mu,h_0,\gamma_0>0$. Then to the leading order of $\mu$, we have $\epsilon_- = -2\sqrt{2}h_0\mu\gamma_0,$ $\epsilon_+=2\sqrt{2}\mu^5/27h_0\gamma_0$. Since $\epsilon_-$ is linear in $\mu$, $E_i$ shows a cube-root dependence on $\mu$ if $h_0\ne0$. The numerical results presented in Fig.\ref{Cubic dispersion} also show the cube-root dispersion near different points at the EL3.
	\section{S3. The EP3 protected by pseudo-chirality}
\begin{figure}
		\centering
		
		\includegraphics[width=1.0\columnwidth]{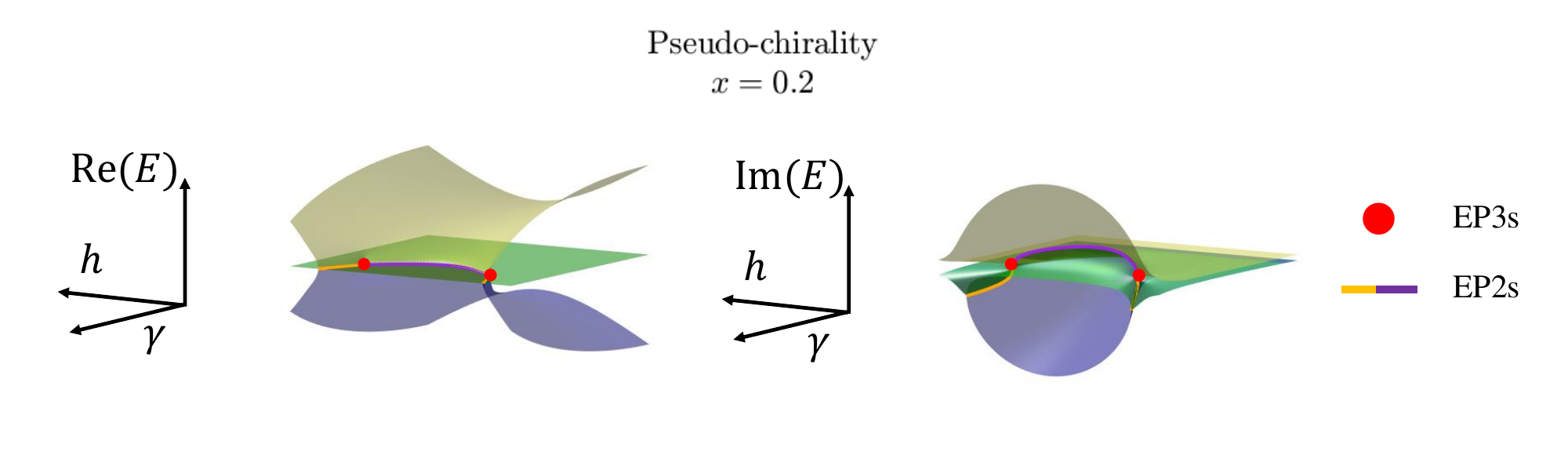}
		\caption{The eigenvalue Riemann sheets of the NH Hamiltonian having pseudo-chirality only.
		}
		\label{psCh energy}
	\end{figure}

	In the main text we have mentioned that when the system satisfies only PT symmetry, isolated EP3s can exist in the 2D parameter space. When the system has only pseudo-chirality, a similar result can be derived. To demonstrate this feature, we introduce another perturbation term to the model Hamiltonian as:
	\begin{equation}
		H^{x}(\gamma,h)=S_x+i\gamma S_z+hS_y+x \left[\begin{array}{ccc}
			0 & 1 & 0\\
			1 & -i & 0\\
			0 & 0 & 0
		\end{array}\right].
	\end{equation}
	It can be verified that when $x\ne0$, PT symmetry is broken and pseudo-chirality is preserved. FIG.\ref{psCh energy} shows the energy surfaces for this Hamiltonian when $x = 0.2$. There are two isolated EP3s, which is consistent with previous discussions.

	\section{S4. The dilation method and the experimental realization in NV center system}
	In this section we show details of experimental realization in NV center.
	
	Fig.~\ref{Pulse_Seq} shows the quantum circuit of the experiment. It contains the state preparation, state evolution and measurement parts. The detailed procedures are as follows.
	
	$\textbf{Step1}:$ The electron spin in NV center, which is a three-level-system, is taken as the system and the nuclear spin is utilized as the ancilla qubit. The static magnetic field is set to 500 Gauss in order to polarize the NV center to state $|0\rangle_e|1\rangle_n$ by optical pumping\cite{PRL_Jacques}.
	After the polarization, the initial state of the coupled system is prepared to the form $|0\rangle_e\otimes(|-\rangle_n+\eta_0|+\rangle_n)$ (not normalized for convenience). Here, $\eta_0$ is properly chosen for the convenience of experimental realization, and $|\pm\rangle_n$ are the eigenstates of $\sigma_y$. In our experiment we have chosen $\eta_0 = \sqrt{0.3}$.
	The preparation is realized by the single-qubit rotation $R_{\phi}(\pi/2)$ on the nuclear spin, where the rotation axis lies in the $x$-$y$ plane and $\phi = \left\lbrace \text{atan}\left[ 2\eta_0/(\eta_0^2-1) \right]+\pi \right\rbrace $ is the angle between the rotation axis and the $x$ axis.
	This nuclear spin rotation is realized by an RF pulse with Rabi frequency calibrated to 5 kHz.
	Then the operation $U_{\rm prep}$ is applied on the electron spin so that the initial state $|\Psi(0)\rangle=(U_{\rm prep}|0\rangle_e)\otimes(|-\rangle_n+\eta_0|+\rangle_n)$ is prepared.
	
	$\textbf{Step2}:$ The evolution under the NH Hamiltonian $H^{(\mu,\nu)}(\gamma,h)$ is realized by constructing $H_{\rm tot}(t)$ based on the dilation method. To be precise, the evolution under $H_{\rm tot}(t)$ gives $|\Psi(t)\rangle=|\psi(t)\rangle_e\otimes|-\rangle_n+\eta(t)|\psi(t)\rangle_e\otimes|+\rangle_n$, where $|\psi(t)\rangle_e$ satisfies the Schr$\ddot{\mathrm{o}}$dinger equation $i\partial_t|\psi(t)\rangle_e=H^{(\mu,\nu)}(\gamma,h)|\psi(t)\rangle_e$. The non-unitary evolution of $|\psi(t)\rangle_e$ can thus be obtained by projecting the dilated state into the subspace where the ancilla qubit state is $|-\rangle_n$. The construction of $H_{\rm tot}(t)$ will be introduced as follows. For simplicity, we drop the subscript $e$ and $n$.
	
	The dilated Hamiltonian takes the form
	\begin{equation}
		H_{\rm tot}(t)=\Gamma(t)\otimes\ket{1}\bra{1}+\Lambda(t)\otimes\ket{0}\bra{0},
	\end{equation}
	where
	\begin{equation}
		\begin{aligned}
			&\Gamma(t)=\left[\begin{array}{ccc}
				d_1(t) & a_1(t) & c_1(t)\\
				a_1^*(t) & d_2(t) & b_1(t)\\
				c_1^*(t) & b_1^*(t) & d_3(t)
			\end{array}\right],\\
			&\Lambda(t)=\left[\begin{array}{ccc}
				d_4(t) & a_2(t) & c_2(t)\\
				a_2^*(t) & d_5(t) & b_2(t)\\
				c_2^*(t) & b_2^*(t) & d_6(t)
			\end{array}\right].
		\end{aligned}
	\end{equation}
	Here $\Gamma(t)$ and $\Lambda(t)$ are complex-valued parameters, and $d_j(t)~(j=1,...,6)$ are real-valued parameters.
	All of the parameters depending on the $H^{(\mu,\nu)}(\gamma,h)$ are obtained by the dilation method. Explicitly, $\Gamma=\hat{\Lambda}+\hat{\Gamma},\Lambda=\hat{\Lambda}-\hat{\Gamma}$, where
	\begin{equation}
		\hat{\Lambda}=[H+(i\partial_t\eta+\eta H)\eta] M^{-1}, \hat{\Gamma} = i( H\eta-\eta H-i\partial_t\eta) M^{-1}.
	\end{equation}
	For simplicity, we have dropped the $t$ dependence and abbreviated $H^{(\mu,\nu)}(\gamma,h)$ as $H$. The explicit forms of $M$ and $\eta$ are
	\begin{equation}
		M(t) = e^{-iH^\dagger t}M(0)e^{iHt},\eta(t)=U(t)[M(t)-I]^{1/2}.
	\end{equation}
	The choice of $M(0)$ and $U(t)$ is flexible. The details can be found in \cite{Dilation}. In order to facilitate the realization of $H_{\rm tot}$ in NV center, we have chosen $M(0) = (\eta_0^2+1)I,U(t)=I$. And $\hat{H}_{\rm tot} = s H_{\rm tot}$ is considered, where $s$ is a fixed nonzero coefficient. This is equivalent to apply the dilation method to the NH Hamiltonian $sH^{(\mu,\nu)}(\gamma,h)$ instead of $H^{(\mu,\nu)}(\gamma,h)$. For simplicity we still use the symbol $H_{\rm tot}$ instead of $\hat{H}_{\rm tot}$ in the following. Then we focus on the realization of $H_{\rm tot}$ in NV center.
	
	$\textbf{Step3}:$ $H_{\rm tot}$ is implemented in NV center by implementing a control Hamiltonian and an appropriate interaction picture. The control Hamiltonian takes the following form (the integral variable $\tau$ is omitted)
	\begin{equation}
		\begin{aligned}
			H_C(t) &= 2\sqrt{2}\pi\Omega_{\rm MW2} \cos[\int_{0}^{t}\omega_{\rm MW2}+\phi_{\rm MW2}(t)]S_x\otimes\ket{1}\bra{1}\\
			&+2\sqrt{2}\pi\Omega_{\rm MW1}  \cos[\int_{0}^{t}\omega_{\rm MW1}+\phi_{\rm MW1}(t)]S_x\otimes\ket{1}\bra{1}\\
			&+2\pi\Omega_{\rm EF1}  \cos[\int_{0}^{t}\omega_{\rm EF1}+\phi_{\rm EF1}(t)]S_{13}\otimes\ket{1}\bra{1}\\
			&+2\sqrt{2}\pi\Omega_{\rm MW4}  \cos[\int_{0}^{t}\omega_{\rm MW4}+\phi_{\rm MW4}(t)]S_x\otimes\ket{0}\bra{0}\\
			&+2\sqrt{2}\pi\Omega_{\rm MW3}  \cos[\int_{0}^{t}\omega_{\rm MW3}+\phi_{\rm MW3}(t)]S_x\otimes\ket{0}\bra{0}\\
			&+2\pi\Omega_{\rm EF2}  \cos[\int_{0}^{t}\omega_{\rm EF2}+\phi_{\rm EF2}(t)]S_{13}\otimes\ket{0}\bra{0}.
		\end{aligned}
	\end{equation}
	The interaction picture is chosen as
	\begin{equation}
		U_{\rm rot} = \exp[i\int_{0}^{t}H_{\rm NV}-\text{diag}(d_1,d_2,\dots,d_6)],
	\end{equation}
	where
	\begin{equation}
		S_x = \frac{1}{\sqrt{2}}\left[\begin{array}{ccc}
			0 & 1 & 0\\
			1 & 0 & 1\\
			0 & 1 & 0
		\end{array}\right], S_{13} = \left[\begin{array}{ccc}
			0 & 0 & 1\\
			0 & 0 & 0\\
			1 & 0 & 0
		\end{array}\right].
	\end{equation}
	Then, the Hamiltonian under the rotating wave approximation can be written as
	\begin{equation}
		\begin{aligned}
			&H_{\rm rot}=U_{\rm rot}H_CU_{\rm rot}^\dagger+\text{diag}(d_1,d_2,\dots,d_6)\\
			&=\left[\begin{array}{ccc}
				d_1 & A_1 & C_1\\
				A_1^* & d_2 & B_1\\
				C_1^* & B_1^* & d_3
			\end{array}\right]\otimes\ket{1}\bra{1}+\left[\begin{array}{ccc}
				d_4 & A_2 & C_2\\
				A_2^* & d_5 & B_2\\
				C_2^* & B_2^* & d_6
			\end{array}\right]\otimes\ket{0}\bra{0},
		\end{aligned}
	\end{equation}
	where
	\begin{equation}
		\begin{aligned}
			&A_1=\pi\Omega_{\rm MW2} e^{-i\phi_{\rm MW2}-i(\int_{0}^{t}\omega_{\rm MW2}-\omega_{12}-d_2+d_1)},\\
			&B_1=\pi\Omega_{\rm MW1} e^{i\phi_{\rm MW1}+i(\int_{0}^{t}\omega_{\rm MW1}-\omega_{23}-d_2+d_3)},\\
			&C_1=\pi\Omega_{\rm EF1} e^{-i\phi_{\rm EF1}-i(\int_{0}^{t}\omega_{\rm EF1}-\omega_{13}-d_3+d_1)},\\
			&A_2=\pi\Omega_{\rm MW4} e^{-i\phi_{\rm MW4}-i(\int_{0}^{t}\omega_{\rm MW4}-\omega_{45}-d_5+d_4)},\\
			&B_2=\pi\Omega_{\rm MW3} e^{i\phi_{\rm MW3}+i(\int_{0}^{t}\omega_{\rm MW3}-\omega_{56}-d_5+d_6)},\\
			&C_2=\pi\Omega_{\rm EF2} e^{-i\phi_{\rm EF2}-i(\int_{0}^{t}\omega_{\rm EF2}-\omega_{46}-d_6+d_4)}.\\
		\end{aligned}
	\end{equation}
	Here we label the energy levels $\ket{m_S=1,0,-1}_e\otimes\ket{m_I=1}_n$ ($\ket{m_S=1,0,-1}_e\otimes\ket{m_I=0}_n$) as 1,2 and 3 (4,5 and 6) for simplicity and $\omega_{ij}$ $(\omega_{ij}>0)$ is the transition frequency between the levels $i$ and $j$.
	Comparing $H_{\rm rot}$ and $H_{\rm tot}$, we obtain the following conditions for the control Hamiltonian:
	\begin{equation}
		\begin{aligned}
			&\omega_{\rm MW2} = \omega_{12}+d_2-d_1,a_1=\pi\Omega_{\rm MW2}e^{-i\phi_{\rm MW2}},\\
			&\omega_{\rm MW1} = \omega_{23}+d_2-d_3,b_1=\pi\Omega_{\rm MW1}e^{i\phi_{\rm MW1}},\\
			&\omega_{\rm EF1} = \omega_{13}+d_3-d_1,c_1=\pi\Omega_{\rm EF1}e^{-i\phi_{\rm EF1}},\\
			&\omega_{\rm MW4} = \omega_{45}+d_5-d_4,a_2=\pi\Omega_{\rm MW4}e^{-i\phi_{\rm MW4}},\\
			&\omega_{\rm MW3} = \omega_{56}+d_5-d_6,b_2=\pi\Omega_{\rm MW3}e^{i\phi_{\rm MW3}},\\
			&\omega_{\rm EF2} = \omega_{46}+d_6-d_4,c_2=\pi\Omega_{\rm EF2}e^{-i\phi_{\rm EF2}}.\\
		\end{aligned}
		\label{solve}
	\end{equation}
	Then the amplitudes $\Omega_{\alpha}(t)$, angular frequencies $\omega_{\alpha}(t)$ and phases $\phi_{\alpha}(t)$ for the six control pulses can be solved from Eq.\ref{solve}, where $\alpha\in\lbrace\rm MW1,MW2,MW3,MW4,EF1,EF2\rbrace$. For example, FIG.2c,d show $\Omega_{\alpha}(t)$ and $\phi_{\alpha}(t)$ for the parameter configurations $h=\mu=\nu=0,\gamma=0.5$ and $h=\mu=\nu=0,\gamma=1.0$, where $s = 2\pi\times40$ kHz is chosen. When $\gamma=0.5$, both the amplitudes $\Omega_{\alpha}$ and phases $\phi_{\alpha}$ are oscillatory, indicating that the eigenvalues are real and the evolution is periodic. And when $\gamma=1.0$, i.e., at the exceptional point, $\Omega_{\alpha}$ and $\phi_{\alpha}$ are no longer oscillatory. The state evolves to a steady state. This is compatible with the results shown in the main text.
	
\begin{figure*}[http]
		\centering
		\includegraphics[width=1\columnwidth]{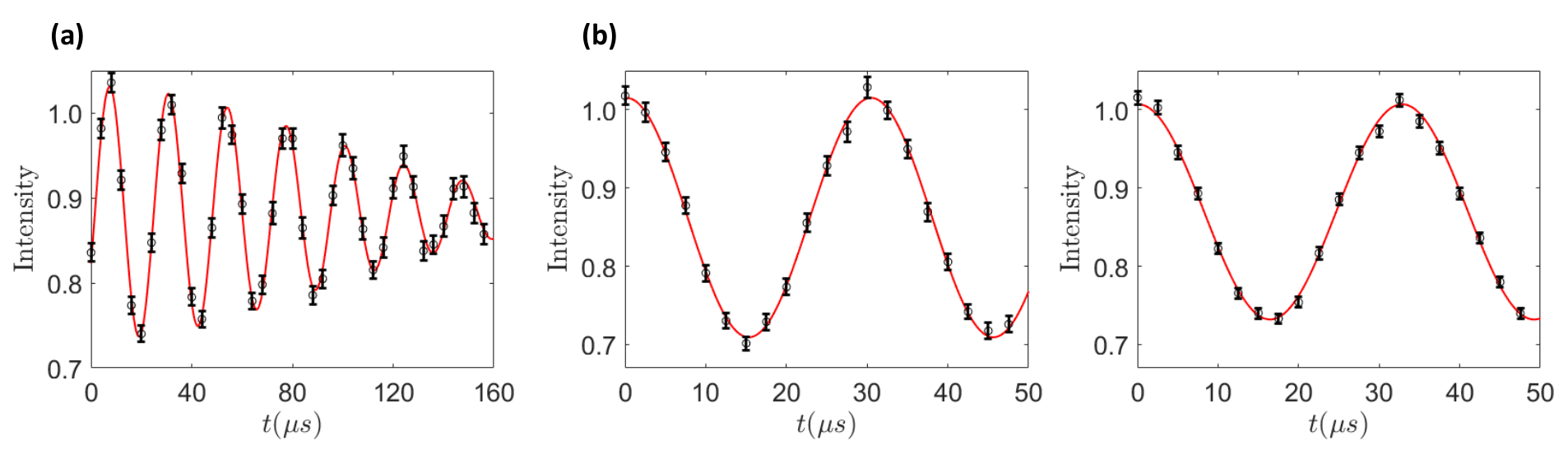}
		\caption{(a) The result of the Ramsey experiment. Fitting the curve gives that $T_2^\star = 127(4)$ $\mathrm{\mu s}$. (b) The Rabi oscillation driven by the AC electric fields. Fitting the curve gives that the Rabi frequencies are $33.0(2)$ kHz and $30.5(2)$ kHz.
		}
		\label{T2 Rabi}
	\end{figure*}

\begin{figure*}[http]
		\centering
		\includegraphics[width=1\columnwidth]{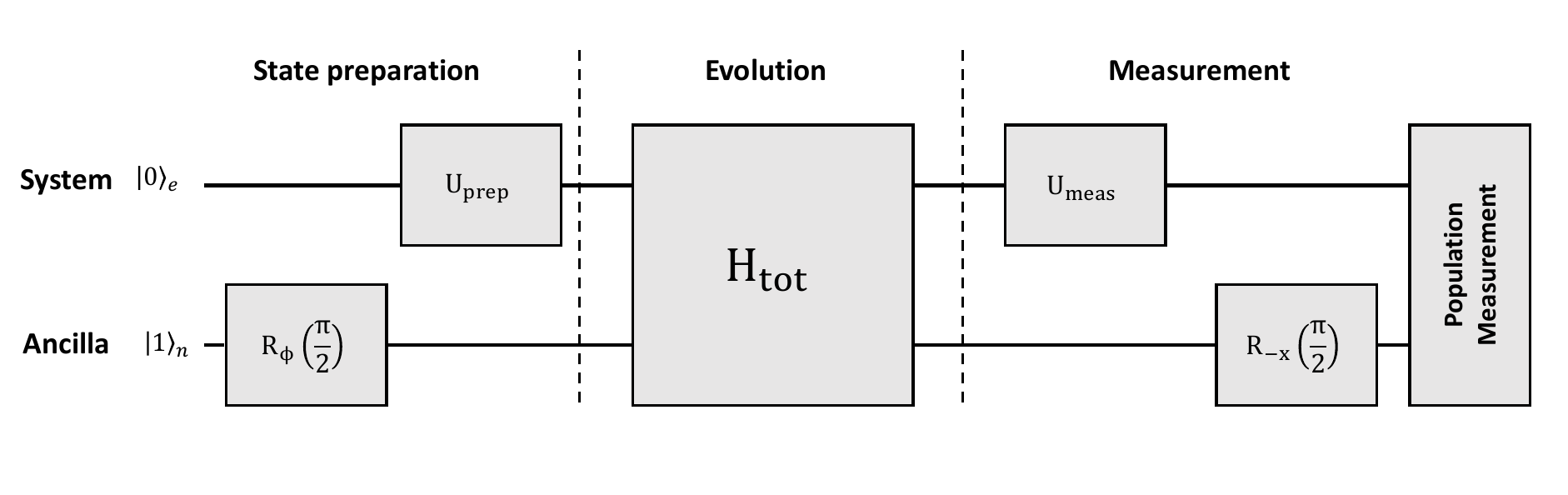}
		\caption{Quantum circuit of the experiment. The electron spin, which is a three-level-system, is taken as the system and the nuclear spin is utilized as the ancilla qubit. $R_{\phi}(\pi/2)$ is the single-qubit rotation on the nuclear spin, where the rotation axis lies in the XY plane and $\phi$ is the angle between the rotation axis and the X axis. After $R_{\phi}(\pi/2)$, $U_{\rm prep}$ is applied to prepare the initial state. Then the coupled system evolves under the dilated Hamiltonian $H_{\rm tot}$. Then $U_{\rm meas}$ is applied to change the measurement basis. Finally the populations of the six energy levels are measured after the rotation $R_{-x}(\pi/2)$.
		}
		\label{Pulse_Seq}
	\end{figure*}

	The off-diagonal elements $a_1(t)$, $b_1(t)$, $a_2(t)$ and $b_2(t)$ of $H_{\rm tot}(t)$ are realized by four selective microwave (MW) pulses.
	The alternating current (AC) electric field is generated by the fabricated electrodes on the diamond surface to realize the off-diagonal elements $c_1(t)$ and $c_2(t)$.
	In our setup the Rabi frequency depends on the voltage applied to the transmission line or electrodes linearly. Thus we calibrate the slope and intercept for MW and electric field pulses. Then the waveforms can be designed according to $\Omega_{\alpha},\phi_{\alpha},\omega_{\alpha}$. The pulses are all generated by an arbitrary waveform generator. The dephasing time of the electron spin of the NV center in our experiments is measured to be $T_2^\star = 127(4)$ $\mathrm{\upmu s}$ (Fig.~\ref{T2 Rabi}a), which is long enough comparing to the experimental evolution time of the order of 30 $\mathrm{\upmu s}$. To characterize the transitions between $|1\rangle_e$ and $|-1\rangle_e$, the experiment results for the Rabi oscillation between $|1\rangle_e|1\rangle_n$ and $|-1\rangle_e|1\rangle_n$ as well as $|1\rangle_e|0\rangle_n$ and $|-1\rangle_e|0\rangle_n$ is shown in Fig.~\ref{T2 Rabi}(b).
	Fitting the photoluminescence intensities yield the Rabi frequencies of $33.0(2)$ kHz and $30.5(2)$ kHz corresponding to the nuclear spin state $|1\rangle_n$ and $|0\rangle_n$, respectively. After the evolution governed by $H_{\rm tot}(t)$, $U_{\rm meas}$ is applied to change the measurement basis.
	
	$\textbf{Step4}:$ Project the dilated state evolution into the subspace where the ancilla is $\ket{-}_n$. To do this, first the nuclear spin rotation $R_{-x}(\pi/2)$ is applied to transform the state $|\Psi(t)\rangle = U_{\rm meas}|\psi(t)\rangle_e |-\rangle_n + U_{\rm meas}\eta(t)|\psi(t)\rangle_e |+\rangle_n$ into $|\Phi(t)\rangle = U_{\rm meas}|\psi(t)\rangle_e |1\rangle_n + U_{\rm meas}\eta(t)|\psi(t)\rangle_e |0\rangle_n$. Then the population evolutions of all six levels are obtained by both the normalization and measurement sequences (the detail will be shown in section S5). The normalization is implemented in the target subspace.
	The normalized population of the electron spin state $\ket{2}$ (denoted as $P_0$), when nuclear spin state is $|1\rangle_n$, is obtained by $P_0=P_{\ket{2}}/(P_{\ket{1}}+P_{\ket{2}}+P_{\ket{3}})$, with $P_{\ket{1}}$, $P_{\ket{2}}$ and $P_{\ket{3}}$ being the population of levels labeled by  $1$, $2$ and $3$, respectively.
	
	In our experimental setup, the RF coil hosts two resonant frequencies $\omega_{25},\omega_{36}$, which drive the transition between $\ket{2}$ and $\ket{5}$, $\ket{3}$ and $\ket{6}$, respectively. In order to rotate the nuclear spin state in the $\ket{1,4}$ subspace in the final $R_{-x}(\pi/2)$ part, we sequentially apply the operations $R_{-x}^{25}(\pi/2)+R_{-x}^{36}(\pi/2)$, $R^{12}(\pi)+R^{45}(\pi)$ and $R_{-x}^{25}(\pi/2)$, where the superscripts mark the transition levels.

	\section{S5. Experimental acquisition of population evolution}
	In this section we introduce the normalization procedures in our experiment and the acquisition of populations from the experimental data. In this section, $L_i$ ($i=1,2,...,6$) stands for the photoluminescence rate for the level labeled by $i$, and $L$ is the column vector with its elements equal to $L_i$. And $S$ is the polarization for the electron spin.
	
	Since the electron spin state is not ideally polarized after the initial laser polarization procedure, we first characterize the polarization of the electron spin. The polarization of the electron spin is measured by a set of independent normalization sequences, where we focus on the subspace spanned by $\ket{2},\ket{3},\ket{5},\ket{6}$. We choose five different sequences (after the laser polarization) to obtain a set of equations for $L_i$ and $S$. The five sequences are chosen to be $\mathbb{I}$, $R^{23}(\pi)$, $R^{25}(\pi)$, $R^{23}(\pi)+R^{36}(\pi)$ and $R^{23}(\pi)+R^{25}(\pi)$. Here $\mathbb{I}$ means the identity operation, i.e., no pulse is applied. We can obtain the set of equations as
	\begin{equation}
		\begin{aligned}
			&L_2S+L_3(1-S)/2 = C^{\rm Pol}_1,\\
			&L_3S+L_2(1-S)/2 = C^{\rm Pol}_2,\\
			&L_5S+L_3(1-S)/2 = C^{\rm Pol}_3,\\
			&L_6S+L_2(1-S)/2 = C^{\rm Pol}_4,\\
			&L_3S+L_5(1-S)/2 = C^{\rm Pol}_5,\\
		\end{aligned}
	\end{equation}
	where $C^{\rm Pol}_j$ is the measured counts for sequence $j$. And the polarization $S$ can be solved from these equations. In our experiment, the result is $S=0.98(1)$.

	In order to extract the population information in the system subspace, both the normalization and measurement sequences are implemented.
	The normalization sequences are utilized to obtain the photoluminescence rate $L_i$ ($i=1,2,...,6$) for each level. After the initial laser pulse, the state of the total system is $\rho_{\rm ini} = (1-S)/2\ket{1}\bra{1}+S\ket{2}\bra{2}+(1-S)/2\ket{3}\bra{3}$. 
	By combining different MW and RF $\pi$ pulses, the normalization sequences rotate $\rho_{\rm ini}$ to different states with the form $\rho = (1-S)/2\ket{i}\bra{i}+S\ket{j}\bra{j}+(1-S)/2\ket{k}\bra{k}$ ($i\ne j,j\ne k,i\ne k, 1\le i,j,k\le 6$).
	Here we use six normalization sequences to obtain a set of equations $L_{i_l}(1-S)/2 +L_{j_l}S+L_{k_l}(1-S)/2 = C_l^{\rm norm}$, where $l = 1,2,...,6$ labels the normalization sequences and $C_l^{\rm norm}$ is the measured counts for normalization sequences. Here the polarization $S$ is taken as the value obtained by the independent polarization measurement. Finally by solving these equations, $L_i$ can be obtained. The normalization sequences for our experiment are $\mathbb{I}$, $R^{12}(\pi)$, $R^{23}(\pi)$, $R^{23}(\pi)+R^{36}(\pi)$, $R^{25}(\pi)$ and $R^{25}(\pi)+R^{45}(\pi)$.
	
	The measurement sequences are utilized to solve the population of each level in our experiment. Knowing the photoluminescence rate for each level, the equations for different measurement sequences are $L^T M_m p = C_m^{\rm exp}$ ($m = 1,2,...,5$), where $p$ is the column vector recording the populations of the six levels, $M_m$ is a permutation matrix depending on the combination of the $\pi$ pulses and $C_m^{\rm exp}$ is the measured counts. Here five different measurement sequences are applied such that the equations are linearly independent. We have chosen the five measurement sequences as $\mathbb{I}$, $R^{12}(\pi)$, $R^{23}(\pi)$, $R^{36}(\pi)+R^{23}(\pi)$ and $R^{25}(\pi)$. With an additional constraint $\sum_{i=1}^{6}p_i = 1$, the populations of each level can be solved. From the population normalization in the subspace spanned by $\ket{1,2,3}$, we obtain the population evolution under the NH Hamiltonian, i.e., $P_0 = p_2/(p_1+p_2+p_3)$.
	
	\section{S6. The verification of parameters of the NH Hamiltonians and the acquisition of eigenvalues}
	The eigenvalues of the non-Hermitian Hamiltonian are obtained from the experimentally retrieved parameters $\gamma_{exp}$, $h_{exp}$, $\mu_{exp}$ and $\nu_{exp}$ based on the form of non-Hermitian Hamiltonian $H^{(\mu,\nu)}(\gamma,h)$.
	Parameters $\mu_{exp}$ and $\nu_{exp}$ can be independently retrieved from the measurement of the symmetry-related conserved quantities.
	Then, $\gamma_{exp}$ and $h_{exp}$ can be obtained from the population evolutions of two different sets of initial states and measurement bases.
	The detailed procedures are as follows.
	We note that in this section, all the experimentally realized Hamiltonians are $\hat{H} = s H$, where $s$ is the nonzero coefficient as in section S4.
	
	\begin{figure*}[http]
		
		\centering
		
		\includegraphics[width=1.0\columnwidth]{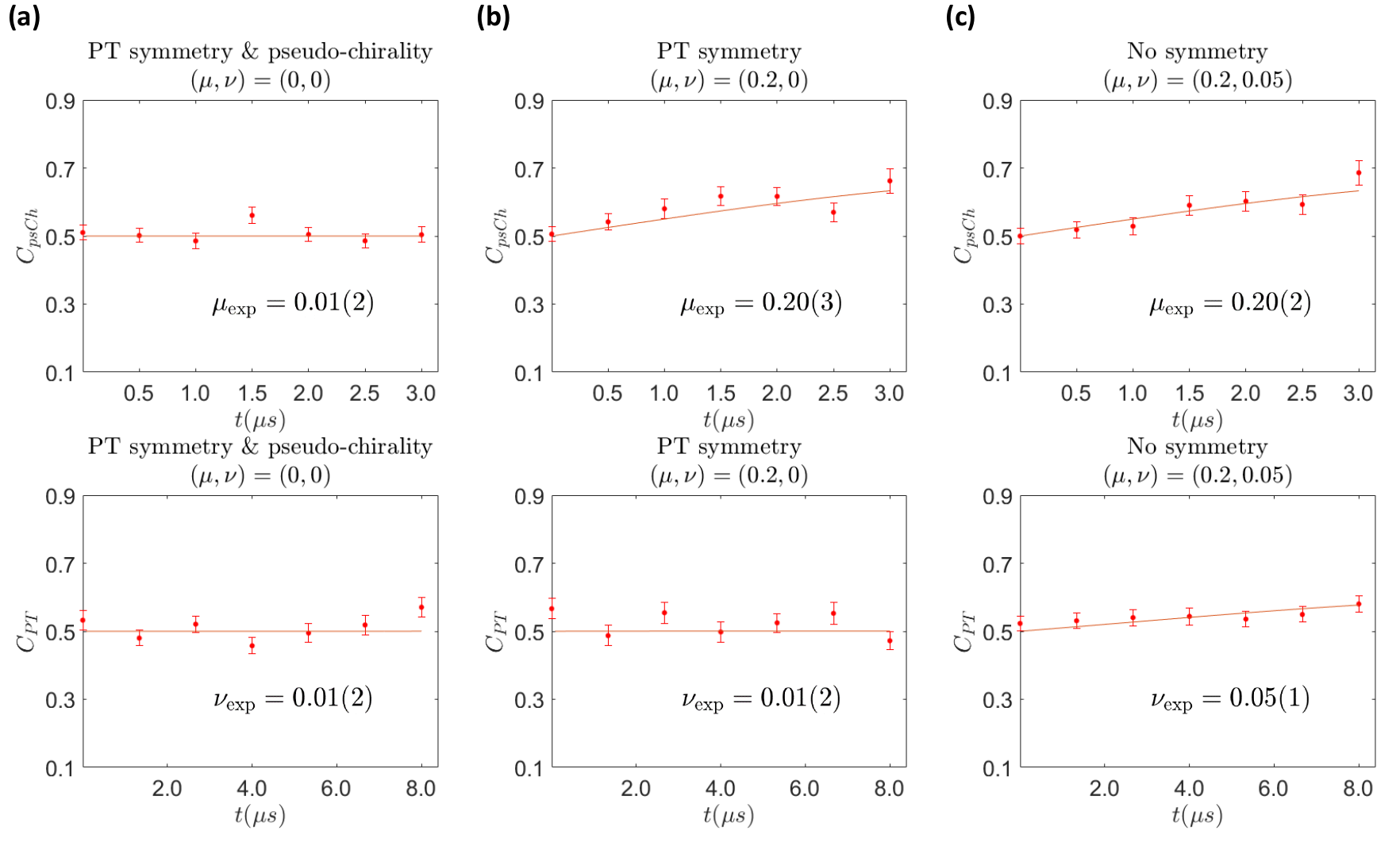}
		\caption{The measured quantities $C_{\rm psCh}$ and $C_{\mathcal{PT}}$ and the retrieved parameters $\mu_{\rm exp},\nu_{\rm exp}$. (a)-(c) are the results for the cases when the system has both symmetries, only $\mathcal{PT}$ symmetry and no specific symmetry, respectively.
		}
		\label{MuNu_exp}
	\end{figure*}
	
	First, by measuring the conserved quantities related to $\mathcal{PT}$ symmetry, we can independently retrieve $\nu$ in the NH Hamiltonian.
		The Hamiltonian $H^{(\mu,\nu)}(\gamma,h)$ in Eq.~1 can be divided into a $\mathcal{PT}$ symmetry-preserving part $H_{\mathcal{PT}}=S_x+i\gamma S_z+hS_y+\mu\left( \begin{array}{ccc}
			0  & 1 & 0\\
			-1 & 0 & -1\\
			0  & 1 & 0
		\end{array}
		\right )$ and a $\mathcal{PT}$ symmetry-breaking part $H_{b}=\nu S_z$.
		The $\mathcal{PT}$ symmetry-preserving part $H_{\mathcal{PT}}$ satisfies the relation $U_{\mathcal{PT}}H_{\mathcal{PT}}U_{\mathcal{PT}}^{-1}=H_{\mathcal{PT}}^*$, where $U_{\mathcal{PT}}$ is shown in Eq.\ref{Unitaries}.
		This leads to  $U_{\mathcal{PT}}e^{-iH_{\mathcal{PT}}t}U_{\mathcal{PT}}^{-1}=e^{-iH_{\mathcal{PT}}^*t}$.
		Define the quantity $C_{\mathcal{PT}}$ as
		\begin{equation}
			C_{\mathcal{PT}}=|\bra{\phi}e^{i\hat{H}^*t}U_{\mathcal{PT}}e^{-i\hat{H}t}\ket{\psi}|^2.
		\end{equation}
		When the time is very short, $C_{\mathcal{PT}}$ takes the following form
		\begin{equation}
			|\bra{\phi}U_{\mathcal{PT}}-ist(U_{\mathcal{PT}}H_{b}-H_{b}^*U_{\mathcal{PT}})+O(t^2)\ket{\psi}|^2.
		\end{equation}
		Here, we choose $\ket{\psi} = (0,(1+i)/2,1/\sqrt{2})^T$ and $\ket{\phi} = ((1+i)/2,1/\sqrt{2},0)^T$. Then we obtain $C_{\mathcal{PT}} = 1/2+s_{1}\nu t+O(t^2)$, where the scaling factor is $s = s_{1} = 2\pi\times30$ kHz. This result also shows that $C_{\mathcal{PT}}$ does not depend on $\gamma,h$ and $\mu$ to the first order of $t$. Therefore, the parameter $\nu$ in the Hamiltonian can be obtained independently by measuring $C_{\mathcal{PT}}$.
		
		\begin{figure}[http]
			
			\centering
			
			\includegraphics[width=0.8\columnwidth]{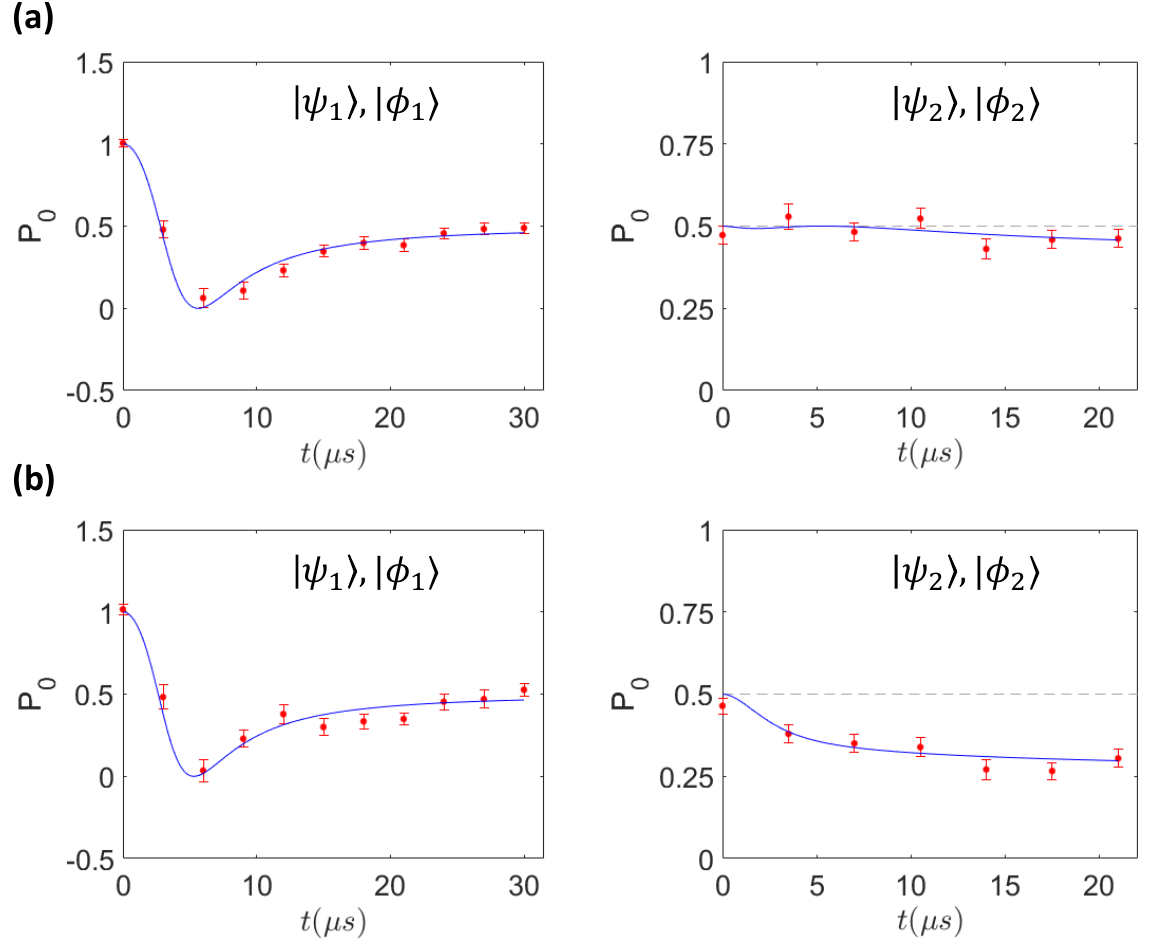}
			\caption{Population evolution under $H^{(\mu,\nu)}(\gamma,h)$ with two different sets of initial states and measurement bases. The parameters $\gamma$ and $h$ are set as $h = 0,\gamma=1$ in (a), and $h = -0.35, \gamma = 1.06$ in (b) (both when $\mu = \nu = 0$).
			}
			\label{h_gamma_exp}
		\end{figure}
		\begin{table*}
			\centering
			\caption{Results of the measured values of $\gamma$ and $h$ for the EP3s on the EL3.}
			\label{hgamma}
			\setlength{\tabcolsep}{2.5mm}{
					\begin{tabular}{c  c c c c c c c}
						\hline
						\hline
						& & & & & & &\\[-6pt]
						$\gamma$&-1.40&-0.75&-0.35&0.00&0.35&0.75&1.40\\
						& & & & & & &\\[-6pt]
						$h$&1.72&1.25&1.06&1.00&1.06&1.25&1.72\\
						& & & & & & &\\[-6pt]
						$\gamma_{\rm exp}$&-1.34(11)&-0.79(11)&-0.34(4)&0.00(3)&0.30(5)&0.77(12)&1.34(11)\\
						& & & & & & &\\[-6pt]
						$h_{\rm exp}$&1.68(16)&1.31(10)&1.06(4)&1.00(2)&1.00(4)&1.27(9)&1.73(11)\\
						\hline
						\hline
				\end{tabular}}
			\end{table*}
			$C_{\mathcal{PT}}$ can be experimentally obtained by sequentially applying the evolution under $s_1H^{(\mu,\nu)}(\gamma,h)$, the unitary operation $U_{\mathcal{PT}}$, the evolution under $-s_1H^{(\mu,\nu)}(\gamma,h)^*$ and the population measurement.
			
			Second, the parameter $\mu$ can similarly be retrieved from
			\begin{equation}
				C_{\rm psCh}=|\bra{\phi}e^{-i\hat{H}^\dagger t}U_{\rm psCh}e^{-i\hat{H}t}\ket{\psi}|^2,
			\end{equation}
			where $U_{\rm psCh}$ makes the Hamiltonian satisfy the condition $U_{\rm psCh}H_{\rm psCh}U_{\rm psCh}^{-1}=-H_{\rm psCh}^{\dag}$ in presence of pseudo-chirality. It can be verified that choosing $\ket{\psi} = (0,1,0)^T$ makes $C_{\rm psCh}$ sensitive to $\mu$ only, up to the first order of $t$. And choosing $\ket{\phi} = (0,1/\sqrt{2},i/\sqrt{2})^T$ gives $C_{\mathcal{\rm psCh}} = 1/2+2s_2\mu t+O(t^2)$, where $s = s_2 = 2\pi\times20$ kHz.
			
			We take the example that the NH Hamiltonian $H^{(\mu,\nu)}(\gamma,h)$ has both $\mathcal{PT}$ symmetry and pseudo-chirality, i.e., when both $\mu$ and $\nu$ are set to 0.
			From the experimental results of the conserved quantities as shown in Fig.~\ref{MuNu_exp}, the values of $\mu$ and $\nu$ are retrieved as 0.01(2) and 0.01(2), respectively. Since the results are insensitive to $\gamma$ and $h$, we have fixed $\gamma=0.7,h=0$ during the measurement of $\mu$ and $\nu$. The experiment results are in good agreement with theoretical predictions, which also shows that the parameters in the NH Hamiltonian can be controlled precisely.

			Third, the parameters $\gamma$ and $h$ are retrieved by measuring the population evolutions under the NH Hamiltonian with two different sets of initial states and measurement bases. In other words, $P_0(t) = |\bra{\phi}e^{-iHt}\ket{\psi}|^2/N(t)$, where $N(t) = |\bra{\psi}e^{iH^\dagger t}e^{-iHt}\ket{\psi}|^2$ is a normalization factor.
			%
			%
			In our experiment, the operation $U_{\rm prep}$ is selected as $U_{\rm prep}=U_{\rm prep1}=I$ and $U_{\rm prep}=U_{\rm prep2}=\left( \begin{array}{ccc}
				1/\sqrt{2}  & -i/\sqrt{2} & 0\\
				-i/\sqrt{2} & 1/\sqrt{2} & 0\\
				0  & 0 & 1
			\end{array}
			\right )$ to prepare $\ket{\psi} = \ket{\psi_1} = (0,1,0)^T$ and $\ket{\psi} = \ket{\psi_2} = (-i/\sqrt{2},1/\sqrt{2},0)^T$, respectively.
			The operation $U_{\rm meas}$ is selected as $U_{\rm meas}=U_{\rm meas1}=I$ and $U_{\rm meas}=U_{\rm meas2}=\left( \begin{array}{ccc}
				1/\sqrt{2}  & -1/\sqrt{2} & 0\\
				1/\sqrt{2} & 1/\sqrt{2} & 0\\
				0  & 0 & 1
			\end{array}
			\right )$ to realize the projection on the states $\ket{\phi} = \ket{\phi_1} = (0,1,0)^T$ and $\ket{\phi} = \ket{\phi_2} = (1/\sqrt{2},1/\sqrt{2},0)^T$, repectively.

			Fig.~\ref{h_gamma_exp} shows two cases of the population evolutions at EP3 when the NH Hamiltonian satisfies both $\mathcal{PT}$ symmetry and pseudo-chirality with $\mu=\nu=0$.
			The parameters $h$ and $\gamma$ are set to $h=0,\gamma=1$ and $h=-0.35,\gamma=1.06$.
			Fitting simultaneously the two population evolutions leads to $h_{\rm exp}=0.00(3),\gamma_{\rm exp}=1.00(3)$ and $h_{\rm exp}=-0.34(4),\gamma_{\rm exp}=1.06(4)$, which are in good agreement with the preset parameters. The other results of measured values of $\gamma$ and $h$ for the EP3s on the EL3 are listed in TABLE. \ref{hgamma}.

			The eigenvalues are evaluated with $(\mu_{\rm exp},\nu_{\rm exp},h_{\rm exp},\gamma_{\rm exp})$ using the model Hamiltonian in Eq.\ref{model}. The errorbars of the eigenvalues are obtained by Monte Carlo method by sampling $(\mu,\nu,h,\gamma)$ using Gaussian distributions according to $(\mu_{\rm exp},\nu_{\rm exp},h_{\rm exp},\gamma_{\rm exp})$ and their variances.

			
			\section{S7. Experimental acquisition of the eigenstates}
			
			In order to verify that the singularity is indeed a 3rd-order exceptional point, we also measured the eigenstates of the NH Hamiltonian.
			\begin{figure}[http]
				
				\centering
				
				\includegraphics[width=1\columnwidth]{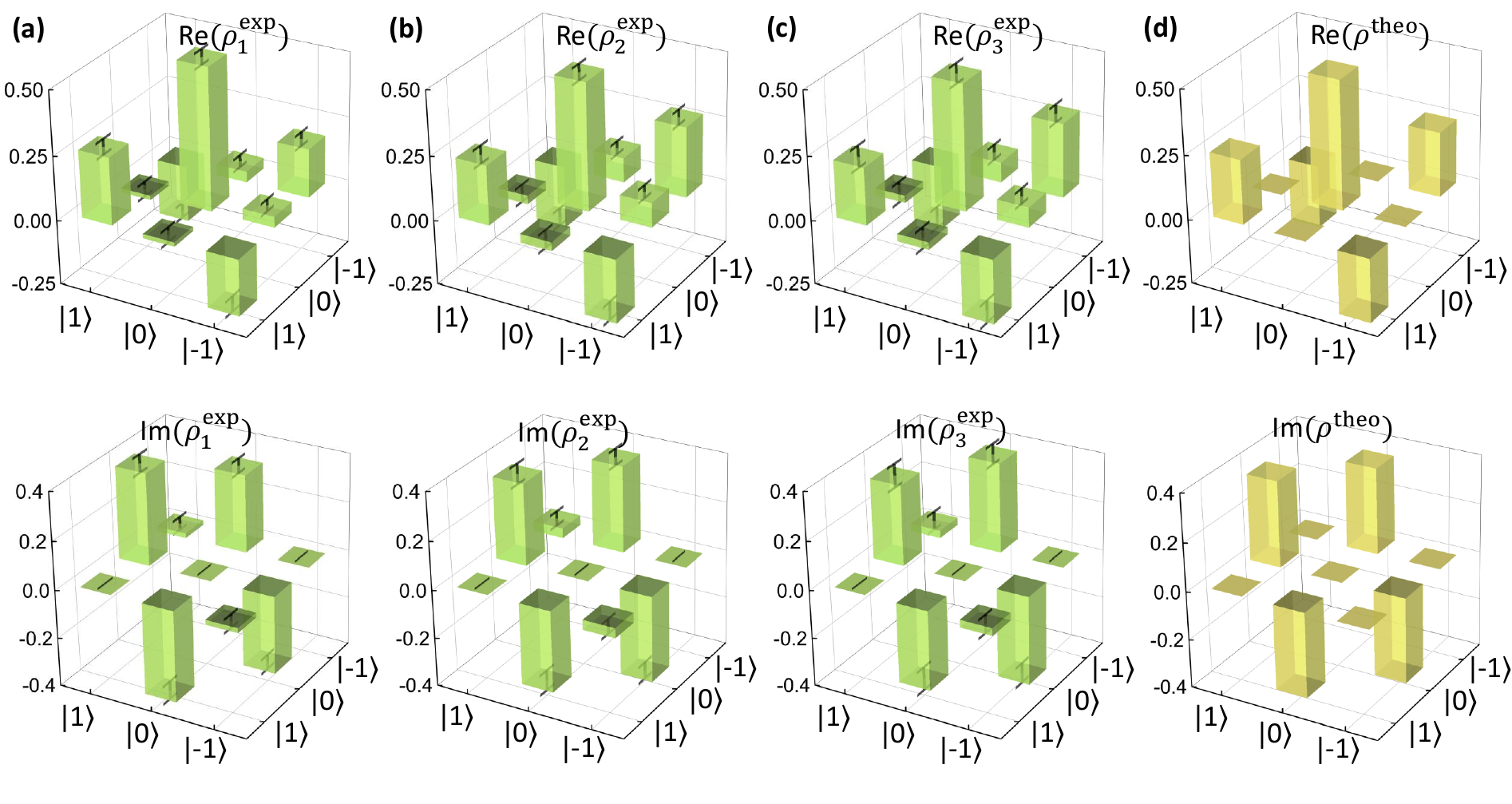}
				\caption{Eigenstates of the NH Hamiltonian at the EP3 where $\mu = \nu=0$, $h = 0$, $\gamma = 1$. (a-c) $\rho_1^{\rm exp}$, $\rho_2^{\rm exp}$ and $\rho_3^{\rm exp}$ are the measured density matrices of three eigenstates (labeled by 1,2 and 3) obtained by quantum state tomography. (d) $\rho_1^{\rm theo}=\rho_2^{\rm theo}=\rho_3^{\rm theo}=\rho^{\rm theo}$ is the density matrix of theoretically predicted eigenstate.
				}
				\label{State EP}
			\end{figure}
			
			\begin{figure}[http]
				
				\centering
				
				\includegraphics[width=1\columnwidth]{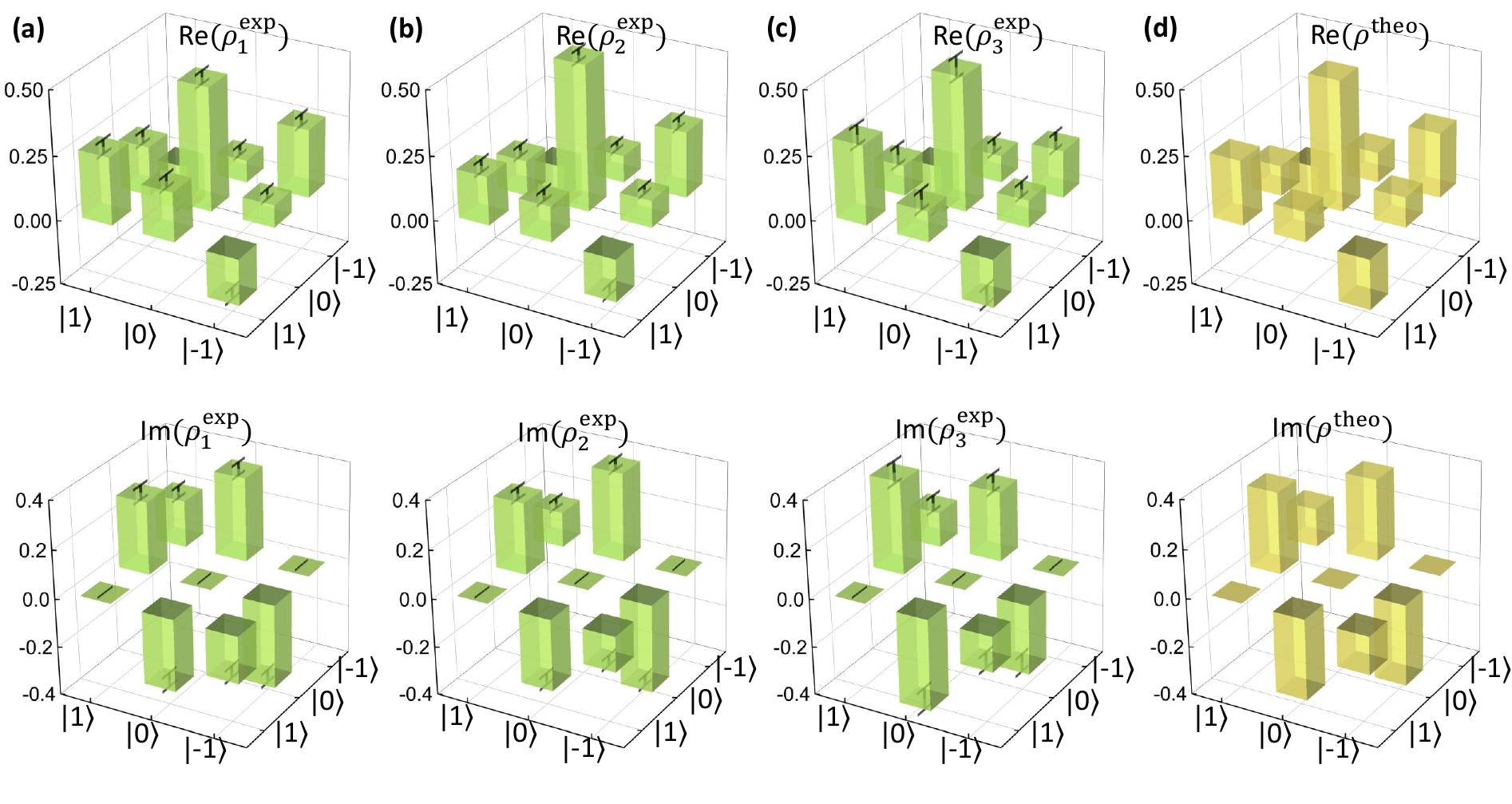}
				\caption{Eigenstates of the NH Hamiltonian at the EP3 where $\mu = \nu=0$, $h = 0.35$, $\gamma = 1.06$. (a-c) $\rho_1^{\rm exp}$, $\rho_2^{\rm exp}$ and $\rho_3^{\rm exp}$ are the measured density matrices of three eigenstates (labeled by 1,2 and 3) obtained by quantum state tomography. (d) $\rho_1^{\rm theo}=\rho_2^{\rm theo}=\rho_3^{\rm theo}=\rho^{\rm theo}$ is the density matrix of theoretically predicted eigenstate.
				}
				\label{State EP3}
			\end{figure}
			
			\begin{figure}[http]
				
				\centering
				
				\includegraphics[width=1\columnwidth]{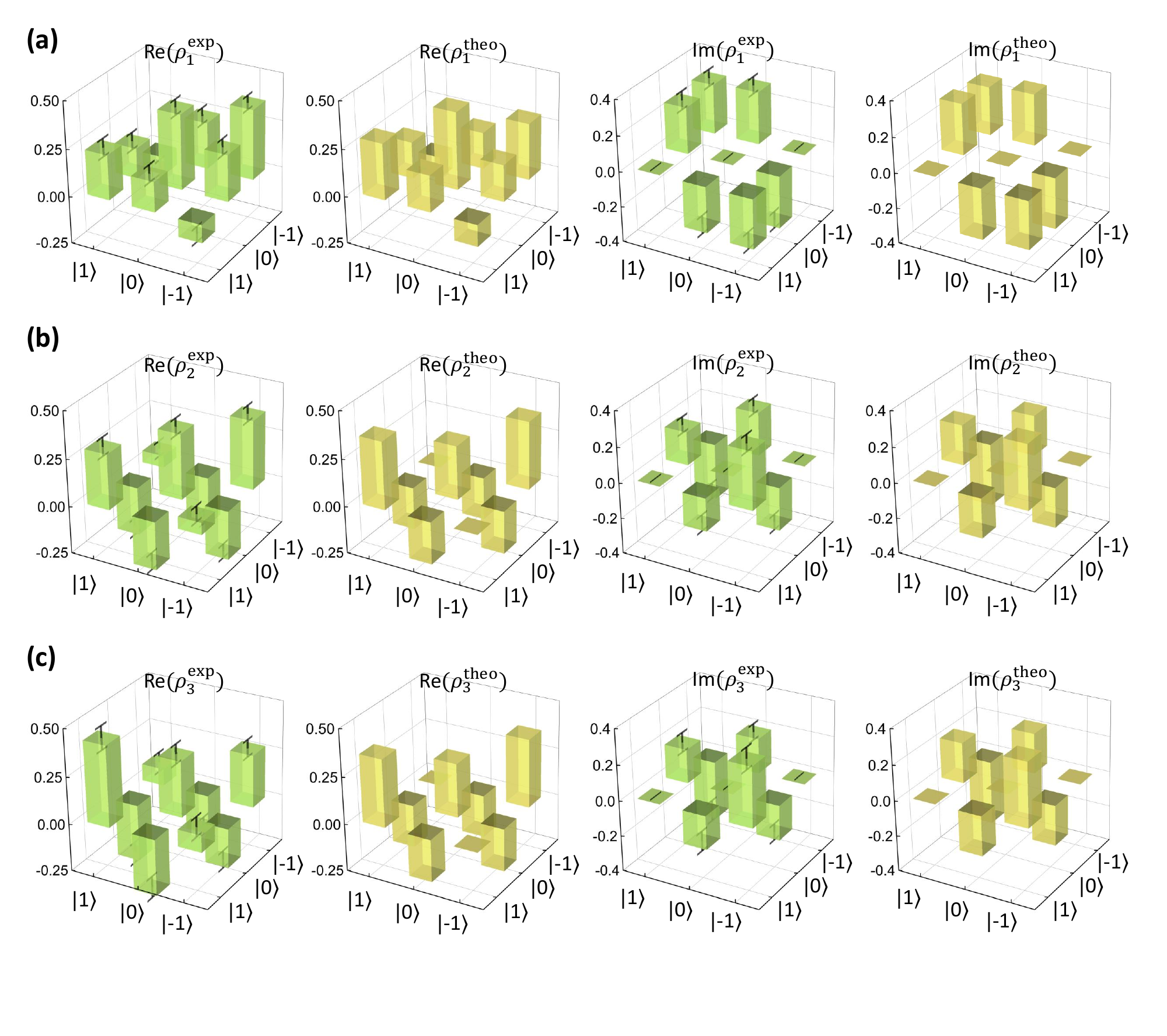}
				\caption{Eigenstates of the NH Hamiltonian at the EP2 where $\mu =0.2$, $\nu=0$, $h = -0.35$, $\gamma = 0.73$. (a-c) $\rho_1^{\rm exp}$, $\rho_2^{\rm exp}$ and $\rho_3^{\rm exp}$ are the measured density matrices of three eigenstates (labeled by 1,2 and 3) obtained by quantum state tomography. $\rho_1^{\rm theo}, \rho_2^{\rm theo}$ and $\rho_3^{\rm theo}$ are the density matrices of theoretically predicted eigenstates. $\rho_2^{\rm theo}$ and $\rho_3^{\rm theo}$ are expected to be degenerate at the EP2.
				}
				\label{State EP2}
			\end{figure}

			The eigenstates of an NH Hamitonian $H$ can be obtained from the steady states under the evolution of the NH Hamiltonian $g(H)$ ($H$ and $g(H)$ have the same eigenstates), where $g(x)$ is an analytic function of $x$. Let $\ket{\psi_{1,2,3}}$ be the eigenstates with complex eigenvalues $E_{1,2,3} = E^r_{1,2,3}+iE^i_{1,2,3}$, where $E^r_n$ and $E^i_n$ are real and imaginary parts of $E_n$, respectively.
			For any initial state written as $\ket{\psi(0)}=c_1\ket{\psi_1}+c_2\ket{\psi_2}+c_3\ket{\psi_3}$, the evolution governed by $g(H)=H$ gives
			\begin{equation}
				\begin{aligned}
					\ket{\psi(t)}&=c_1 e^{-iE^r_1t+E^i_1t}\ket{\psi_1}\\&+c_2 e^{-iE^r_2t+E^i_2t}\ket{\psi_2}\\&+c_3 e^{-iE^r_3t+E^i_3t}\ket{\psi_3}.
				\end{aligned}
			\end{equation}
			Without loss of generality, we assume $E^i_1>E^i_2>E^i_3$.
			The steady state of non-unitary evolution will be the eigenstate $\ket{\psi_1}$ since $E^i_1$ is the largest among $E^i_n$.
			By implementing evolution governed by $g(H)=-H$, the eigenstate $\ket{\psi_3}$ can be obtained, which is due to $-E^i_3>-E^i_2>-E^i_1$.
			To obtain the eigenstate $\ket{\psi_2}$, the evolution is set to be governed by $g(H) = 1/(H-\alpha I)$, where the parameter $\alpha$ is chosen to be closest to $E_2$ such that $\text{Im}(1/(E_2-\alpha))>\text{Im}(1/(E_1-\alpha))$ and $\text{Im}(1/(E_2-\alpha))>\text{Im}(1/(E_3-\alpha))$.
			
			\begin{figure}[http]
				
				\centering
				
				\includegraphics[width=1\columnwidth]{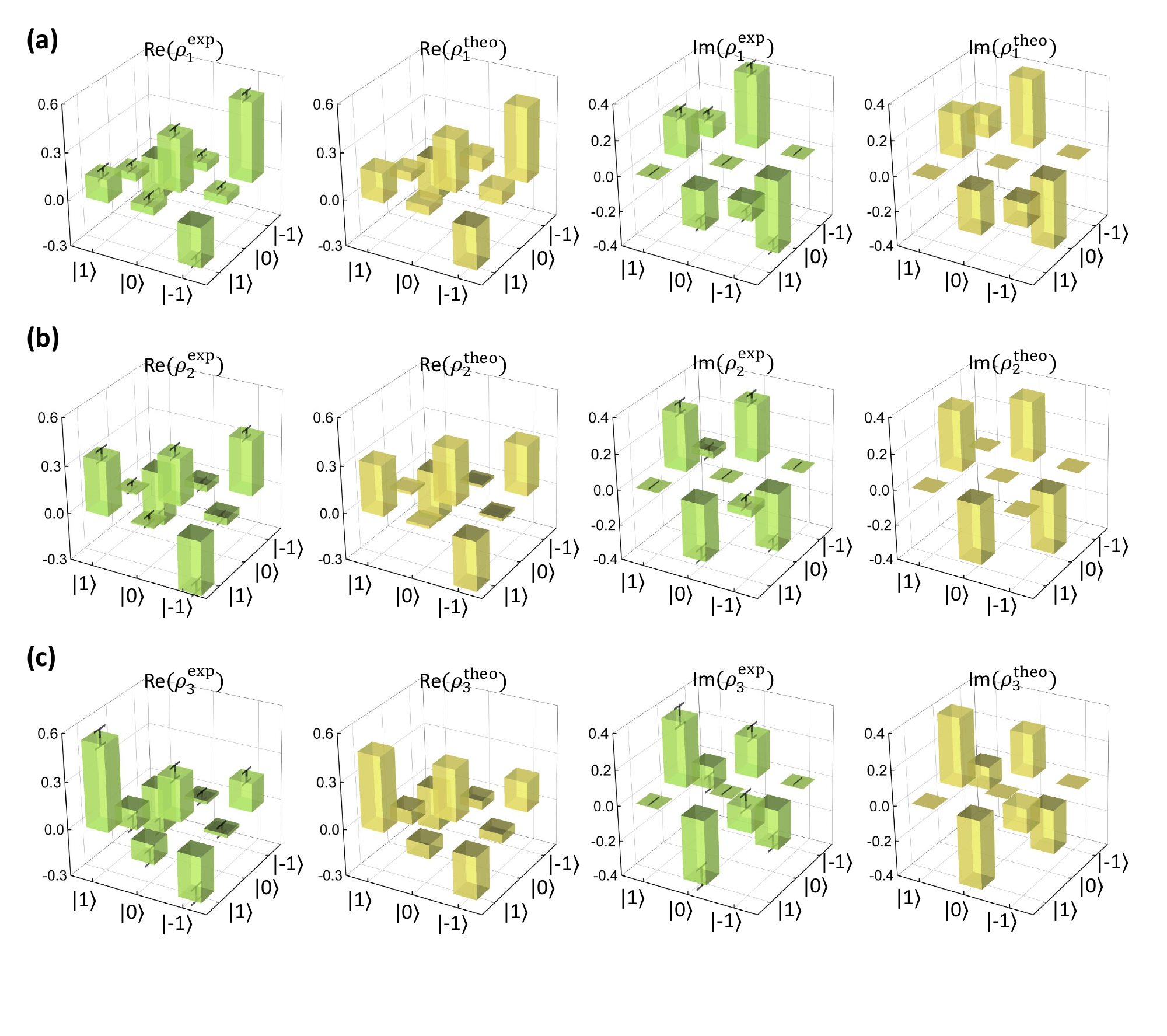}
				\caption{Eigenstates of the NH Hamiltonian with $\mu =0.2$, $\nu=0.05$, $h = 0$, $\gamma = 0.96$. (a-c) $\rho_1^{\rm exp}$, $\rho_2^{\rm exp}$ and $\rho_3^{\rm exp}$ are the measured density matrices of three eigenstates (labeled by 1,2 and 3) obtained by quantum state tomography. $\rho_1^{\rm theo}, \rho_2^{\rm theo}$ and $\rho_3^{\rm theo}$ are the density matrices of theoretically predicted eigenstates.
				}
				\label{State no EP}
			\end{figure}
			
			Here we implemented $g(H) = H,-H$ and $-i/(H-1.5 I)$ to obtain the eigenstates labeled by $1,3$ and $2$ for $H = H^{(\mu,\nu)}(\gamma,h)$, respectively. The eigenstates can then be experimentally obtained by implementing quantum state tomography procedures. The details of the quantum state tomography are shown as follows. Suppose that the final state after the NH evolution is expressed as $\rho = (\rho)_{ij}, i,j=1,2,...,6$. Here the basis states are chosen as $\ket{1,0,-1}_e\otimes\ket{-}_n$ and $\ket{1,0,-1}_e\otimes\ket{+}_n$, corresponding to the labels from 1 to 6, respectively. Under these bases, the system state can be expressed as $\rho_s=(\rho)_{ij}/N,i,j=1,2,3$, where $N = \rho_{11}+\rho_{22}+\rho_{33}$ is a normalization factor. From measurements using the sequences discussed in section S4, the diagonal elements of $\rho$ have already been extracted. $\rho_{12,23,31}$ can be obtained by similar procedures, with appropriately chosen measurement sequences. We take the measurement of $\text{Re}(\rho_{23})$ as an example. The unitary operations are denoted as
			\begin{equation}
				\begin{aligned}
					&U_1 = \left[\begin{array}{ccc}
						1 & 0 & 0\\
						0 & 1/\sqrt{2} & -1/\sqrt{2}\\
						0 & 1/\sqrt{2} & 1/\sqrt{2}
					\end{array}\right]\otimes 1_2, \\&U_2 = \ket{0}\bra{0}\otimes\left[\begin{array}{ccc}
						1/\sqrt{2} & i/\sqrt{2}\\
						i/\sqrt{2} & 1/\sqrt{2}
					\end{array}\right], \\&U_3 = \ket{-1}\bra{-1}\otimes\left[\begin{array}{ccc}
						1/\sqrt{2} & i/\sqrt{2}\\
						i/\sqrt{2} & 1/\sqrt{2}
					\end{array}\right],\\& U_4 = \left[\begin{array}{ccc}
						1 & 0 & 0\\
						0 & 0 & -i\\
						0 & -i & 0
					\end{array}\right]\otimes (1_2+\sigma_z)/2.
				\end{aligned}
			\end{equation}
			These operations can be realized by MW and RF pulses. The total unitary operations are designed as $U_a = U_3U_2U_1$ and $U_b = U_4U_a$. The measured fluorescence for these two cases can be formulated as $C_{a,b} = L^T\cdot$diag$(U_{a,b}\rho U_{a,b}^\dagger)$, where $L$ is the photoluminescence rate vector mentioned in section S4. It can be verified that $C_a-C_b = 2\text{Re}(\rho_{23})(L_3-L_2)$. Hence $\text{Re}(\rho_{23})$ can be obtained. The other elements can be obtained similarly. Since we have five measurement sequences for the diagonal elements and twelve sequences for off-diagonal elements ($\text{Re}(\rho_{23,12,13})$ and $\text{Im}(\rho_{23,12,13})$, each one needs two independent sequences), totally seventeen sequences are used to extract $\rho_s$.

			Since the result directly obtained from above may be a non-physical state, we use the maximum likelihood estimation (MLE) method\cite{PRA_James} to achieve the density matrices of the eigenstates.
			Fig. \ref{State EP}-\ref{State no EP} show the results of the quantum state tomography for the eigenstates of the non-Hermitian Hamiltonian with different parameters.
			The obtained fidelities of the experimental eigenstates with respect to the theoretical ones are higher than 0.98 on average.
			Fig. \ref{State EP} and Fig. \ref{State EP3} show the results of the quantum state tomography for the eigenstates at the EP3 on the EL.
			The overlap between two of the eigenstates is characterized by the fidelity, $F_{ij} = [\mathrm{Tr}(\sqrt{\sqrt{\rho_i}\rho_j\sqrt{\rho_i}})]^2$, between the eigenstates with the eigenvalues $E_i$ and $E_j$.
			For the case of the EP3 at $\mu = \nu=0$, $h = 0$, $\gamma = 1$ as shown in Fig. \ref{State EP}, $F_{13},F_{12}$ and $F_{23}$ are 0.99(2), 0.99(2) and 1.00(2), respectively.
			The eigenstates shown in Fig. \ref{State EP3} at the EP3 where $\mu = \nu=0$, $h = 0.35$, $\gamma = 1.06$ give the fidelities of $F_{13}=0.98(2)$, $F_{12}=0.99(3)$ and $F_{23}=0.98(2)$.
			The coalescence of three eigenstates together with triple degeneracy of eigenvalues shown in Fig. 3b conclusively identifies the EP3s.
			The other fidelities $F_{ij}$ of the eigenstates for the EP3s on the EL are summarized in Table 1 in the main text.
			Furthermore, in order to show the relationship between the structure of EP3 and the symmetry of the non-Hermitian Hamiltonian, we also provide the eigenstate data under the parameter $\mu =0.2$, $\nu=0$, $h = -0.35$, $\gamma = 0.73$ as shown in Fig. \ref{State EP2}.
			In the case where Hamiltonian has only PT symmetry, there is no EP3 when $h=-0.35$.
			However, an EP2 can be observed with $\gamma=0.73$, where only the two eigenstates $\rho_2$ and $\rho_3$ coalesce.
			This result also verifies from the side that when the Hamiltonian only has PT symmetry, there is only isolated EP3.
			When both the PT symmetry and the pseudo-chirality are broken, there is no EP3.
			Fig. \ref{State no EP} shows the eigenstates at $\mu =0.2$, $\nu=0.05$, $h = 0$, $\gamma = 0.96$.
			The three eigenstates are different from each other which are also verified by the fidelities of $F_{13}=0.72(5)$, $F_{12}=0.91(4)$ and $F_{23}=0.92(4)$.
			These results of the eigenstates combined with the corresponding eigenvalues clearly show that there is no EP3 when the non-Hermitian Hamiltonian has no symmetry.
			%
			%
			

			\section{S8. Topological characterization of EL3}
			In this section, we briefly discuss how to realize the topological characterization of the EL3 in our model.
			
			As noted in \cite{PRB_topo,PRB_topo2}, to describe the topology of defects, we can first take a surrounding surface that encloses the defect, then seek for a topological characterization of the surrounding surface. We note that in our context only EPs with order three or higher are considered as defects. Thus in presence of both symmetries, the EP3s form an EL3 and the surrounding surface consists of two points $\lbrace+,-\rbrace$.
			
			Inspired by \cite{PRB_topo,PRB_topo2}, we define the topological invariant as
			\begin{equation}
				W = \frac{1}{2}(\mathrm{sgn}(\mathrm{Re}(R_1(\gamma_+,h_+)))-\mathrm{sgn}(\mathrm{Re}(R_1(\gamma_-,h_-)))),
			\end{equation}
			where the parameters $(\gamma_+,h_+)$ and $(\gamma_-,h_-)$ correspond to the two points $\lbrace+,-\rbrace$, respectively. We find $W = 1$ in our model when the two points are on different sides of the EL3. And $W = 0$ if the two points are on the same side of the EL3.

\end{document}